\pdfoutput=1
\documentclass[11pt,twoside,a4paper,cmspaper,final,collab]{cms-tdr}
\def\svnVersion{b514283}\def\svnDate{2025/11/27}\def\cmsCernNoTag{CERN-EP-2025-005}\def\cmsCernDate{\today}\def\cmsMessage{Published in the European Physical Journal C as \href{http://dx.doi.org/10.1140/epjc/s10052-025-14713-w}{\doi{10.1140/epjc/s10052-025-14713-w}.}}
\begin{document}\cmsNoteHeader{MLG-23-005}

\newlength\cmsTabSkip\setlength{\cmsTabSkip}{1ex}
\newcommand{\HTT}{\ensuremath{\PH\to\PGt\PGt}\xspace}
\newcommand{\Multinomial}{\ensuremath{\mathbb{P}(\{k_{l}\},\,n,\,\{p_{l}\})}\xspace}
\newcommand{\Poisson}{\ensuremath{\mathcal{P}(\,\cdot\,|\,\cdot\,)}\xspace}
\newcommand{\CENNT}{\ensuremath{\text{CENNT}}\xspace}
\newcommand{\SANNT}{\ensuremath{\text{SANNT}}\xspace}
\newcommand{\etau}{\ensuremath{\Pe\tauh}\xspace}
\newcommand{\pp}{\ensuremath{\Pp\Pp}\xspace}
\newcommand{\PePe}{\ensuremath{\Pe\Pe}\xspace}
\newcommand{\PgmPgm}{\ensuremath{\PGm\PGm}\xspace}
\newcommand{\PgtPgt}{\ensuremath{\PGt\PGt}\xspace}
\newcommand{\ZTT}{\ensuremath{\PZ\to\PgtPgt}\xspace}
\newcommand{\ZLL}{\ensuremath{\PZ\to\Pell\Pell}\xspace}
\newcommand{\Wjets}{\ensuremath{\PW}{+}\text{jets}\xspace}
\newcommand{\FF}{\ensuremath{F_{\mathrm{F}}}\xspace}
\newcommand{\qqH}{\ensuremath{\PQq\PQq\PH}\xspace}
\newcommand{\ggH}{\ensuremath{\Pg\Pg\PH}\xspace}
\newcommand{\vx}{\ensuremath{\mathbf{x}}\xspace}
\newcommand{\mjj}{\ensuremath{m_{\mathrm{jj}}}\xspace}
\newcommand{\Njet}{\ensuremath{N_{\text{jet}}}\xspace}
\newcommand{\Nbtag}{\ensuremath{N_{\text{Btag}}}\xspace}
\newcommand{\mtt}{\ensuremath{m_{\PGt\PGt}}\xspace}
\newcommand{\mvis}{\ensuremath{m_\text{vis}}\xspace}
\newcommand{\yl}{\ensuremath{\hat{y}_{l}(\,\cdot\,)}\xspace}
\newcommand{\y}{\ensuremath{\hat{y}(\,\cdot\,)}\xspace}
\newcommand{\rs}{\ensuremath{r_{s}}\xspace}
\newcommand{\Drs}{\ensuremath{\Delta r_{s}}\xspace}
\newcommand{\DrsStat}{\ensuremath{\Delta r_{s}^{\text{stat}}}\xspace}
\newcommand{\LH}{\ensuremath{\mathcal{L}}\xspace}
\newcommand{\Bprop}{\ensuremath{\mathcal{B}_{i}}\xspace}
\newcommand{\Asimov}{\ensuremath{D_{H}^{\mathcal{A}}}\xspace}
\newcommand{\NLL}{\ensuremath{-2\Delta\log\LH}\xspace}
\newcommand{\DataModel}{\ensuremath{\Omega_{X}}\xspace}
\newcommand{\talpha}{\ensuremath{\langle t_{\alpha}\rangle}\xspace}
\newcommand{\SubS}{\ensuremath{\mathcal{S}_{\mathrm{S}}}\xspace}
\newcommand{\SubB}{\ensuremath{\mathcal{S}_{\mathrm{B}}}\xspace}
\newcommand{\DBkappa}{\ensuremath{\mathcal{C}_{\kappa}}\xspace}
\newcommand{\jettau}{\ensuremath{\text{jet}\to\tauh}\xspace}
\newcommand{\LSANNT}{\ensuremath{L_{\SANNT}}\xspace}
\newcommand{\rinc}{\ensuremath{r_{\text{inc}}}\xspace}
\newcommand{\rggH}{\ensuremath{r_{\ggH}}\xspace}
\newcommand{\rqqH}{\ensuremath{r_{\qqH}}\xspace}
\newcommand{\ol}{\ensuremath{\omega_{\lambda}}\xspace}
\newcommand{\CEmin}{\ensuremath{\mathrm{CE'}_{\text{min}}}\xspace}
\newcommand{\etauID}{\ensuremath{\epsilon_{\Pgt}^{\mathrm{ID}}}\xspace}
\newcommand{\etauTrig}{\ensuremath{\epsilon_{\Pgt}^{\text{trg}}}\xspace}
\newcommand{\missIDe}{\ensuremath{\mathrm{ID}_{\Pe}^{\text{miss}}}\xspace}
\newcommand{\thrnd}{\ensuremath{\theta_{\text{rnd}}}\xspace}
\newcommand{\thj}{\ensuremath{\theta_{j}}\xspace}

\cmsNoteHeader{MLG-23-005}
\title{Development of systematic uncertainty-aware neural network trainings for binned-likelihood analyses at the LHC}
\date{\today}

\abstract{
We propose a neural network training method capable of accounting for the effects 
of systematic variations of the data model in the training process and 
describe its extension towards neural network multiclass classification. The 
procedure is evaluated on the realistic case of the measurement of Higgs boson 
production via gluon fusion and vector boson fusion in the $\PGt\PGt$ decay 
channel at the CMS experiment. The neural network output functions are used to 
infer the signal strengths for inclusive production of Higgs bosons as well as 
for their production via gluon fusion and vector boson fusion. We observe 
improvements of 12 and 16\% in the uncertainty in the signal strengths for gluon 
and vector-boson fusion, respectively, compared with a conventional neural network 
training based on cross-entropy.
}
\hypersetup{
pdfauthor={CMS Collaboration},
pdftitle={Development of systematic uncertainty-aware neural network trainings for binned-likelihood analyses at the LHC},
pdfsubject={CMS},
pdfkeywords={CMS, LHC, machine learning, systematic aware, Higgs}}

\maketitle
\section{Introduction}\label{sec:introduction}

A physics measurement implies the synthesis of a potentially huge data set into 
a single or small number of characterizing model parameters of interest (POIs) 
and their uncertainties. The sensitivity of a measurement is determined by its 
power in extracting all information relevant for the determination of the POIs 
from data, such that the most sensitive measurement is the one with the smallest 
uncertainties in the POIs.

Higgs boson (\PH) production at the CERN LHC~\cite{Aad:2012tfa,Chatrchyan:2012ufa,
Chatrchyan:2013lba,CMS:2022dwd,ATLAS:2024fkg} is a rare process, which, depending 
on the \PH decay, may be covered by an overwhelmingly large background from 
standard model (SM) processes not related to \PH physics. This is a particular 
challenge for analyses where the \PH decays into fermions, such as \PQb 
quarks~\cite{ATLAS:2020bhl,CMS:2023vzh} or tau leptons (\HTT)~\cite{CMS:2022kdi,
ATLAS:2024wfv}. At the same time, the complex and feature-rich nature of these 
decay channels makes them ideal candidates for the application of modern 
machine learning (ML) methods, typically exploited to separate signal from 
background processes. The latest measurements of differential simplified template 
cross sections (STXS) from \HTT events of the CMS Collaboration~\cite{CMS:2022kdi} 
form the most precise measurements of this decay channel to date. These measurements 
rely heavily on the use of ML methods in the form of neural network (NN) multiclass 
classification using categorical cross-entropy (CE)~\cite{cross-entropy} as the 
loss function for the NN training. In this paper we refer to this kind of NN 
training as (categorical) CE-based NN training (\CENNT).

If the NN output function can be interpreted as a probability $p_{l}$ to associate 
a given event to class $l$, the categorical CE is equivalent to the negative 
log-likelihood of the multinomial distribution 
\begin{equation}
      \Multinomial = n!
      \prod\limits_{l=1}^{\Lambda} \frac{p_{l}^{k_{l}}}{k_{l}!},
\end{equation}
with
\begin{equation*}
      \sum\limits_{l=1}^{\Lambda}k_{l} = n \qquad \text{and} \qquad
      \sum\limits_{l=1}^{\Lambda}p_{l} = 1, 
\end{equation*}
where $n$ refers to the inclusive number of events, $k_{l}$ to the number of 
events in class $l$, and $\Lambda$ to the number of all classes. Hence, \CENNT 
is a maximum likelihood estimate of a multiclass-classification model based on 
\Multinomial. If successful, it provides the best possible separation of all 
signal and background classes based on the uncertainties related to the statistical 
nature of the measurement, also referred to as aleatoric or statistical 
uncertainties. However, \CENNT is blind to any influence of imperfect knowledge 
of the underlying model \DataModel used to extract a POI related to a given 
measurement from data, where $X$ refers to the set of observables on which the 
measurement is based. We usually quantify our lack of knowledge in terms of 
variations of external parameters of \DataModel resulting in what we call 
epistemic or systematic uncertainties. 

The measurements of Ref.~\cite{CMS:2022kdi} have entered an era of precision where 
the uncertainties, even of differential measurements, are dominated by their systematic 
(rather than statistical) component. This statement holds even if measurements 
will at the same time become increasingly differential, since the more differential 
a measurement becomes the more susceptible it is to imperfect knowledge of external 
model parameters. In this situation, the advantage of using ML methods to maximize the 
sensitivity of a given measurement might be reduced or even lost. It could be 
preserved, however, if an NN training can be designed to take into account not 
only statistical, but also systematic uncertainties. In this paper, we will refer 
to this kind of NN training as systematic uncertainty aware NN training (\SANNT). 
By construction, \SANNT incorporates more information of \DataModel than \CENNT. 
In this sense it is likely to require a larger investment of the analyst in 
characterizing \DataModel. In turn, the presented \SANNT is applicable to any kind 
of binned likelihood analysis and should lead to at least the same or a higher 
sensitivity to a given POI than \CENNT, when based on the same data model. 
Providing an objective strategy for obtaining the most sensitive result for a 
given \DataModel that represents a given understanding of the data, \SANNT thus 
offers a sizable impact on many analyses in the transition from statistically to 
systematically dominated regions, at the LHC and beyond. A number of strategies 
and corresponding loss functions for \SANNT have been proposed in the recent 
literature~\cite{Neal,Cranmer:2015bka,inferno,Wunsch:2020iuh,Simpson:2022suz}. 
These go significantly beyond approaches that aim at only reducing the sensitivity 
of the NN output function to variations of input features in $X$, \eg as introduced 
and discussed in Refs.~\cite{Louppe:2016ylz,Wunsch:2019qbo}, which are usually 
applied to increase the robustness of classification algorithms against differences 
of input features between the training and test data. This work represents the first 
application of a SANNT to an existing LHC analysis that includes a complete and 
realistic model of the systematic uncertainties. It marks an important step towards 
demonstrating the applicability of a SANNT in high-energy particle physics measurements.

{\tolerance=800
In this paper we demonstrate the application of \SANNT to the comprehensive realistic 
data model \DataModel presented in Ref.~\cite{CMS:2022kdi}. Our work is based on 
the method of Ref.~\cite{Wunsch:2020iuh}, replacing CE by the extended binned 
likelihood commonly used for the extraction of a given signal in addition to a 
set of known background processes, based on histogram templates. From this likelihood, 
which can be conceptually extended to include an arbitrary number of nuisance 
parameters related to systematic variations in \DataModel, we derive the uncertainty 
in a given POI through the Fisher information~\cite{fisher1922} and the Cram\'{e}r-Rao 
bound~\cite{cramer1999mathematical,rao1992information}. In this way, we preserve 
maximal congruency of the training target with the measurement target also in the 
presence of systematic uncertainties. Since it intrinsically relies on a predefined 
binning of the output node values through the binned likelihood, the \SANNT described 
here can be viewed as a method to distribute the events optimally across these 
bins to obtain the minimal uncertainty in the POI, during training time. A conventional bin 
optimization procedure in one dimension applied to a \CENNT receives fixed values of 
the NN output function depending on the input space vectors $\vx\in X$, after training. 
Based on these output values the procedure leads to an adjustment of the bin boundaries 
of the histogram to be used for signal extraction. For the \SANNT as described here, 
the binning is fixed, while the adjustment happens to the values of the NN output function, 
during the training. We note that this strictly holds only for tasks where the 
uncertainties in the POI are still dominated by the statistical part, and if a high 
correlation between the NN outputs after \CENNT and \SANNT is retained, \eg by the choice 
of pre-training conditions.
\par}

During training, the minimization of the negative log-likelihood is performed with 
backpropagation. The issue of differentiating discrete histogram boundaries is 
solved through custom functions in the backward pass. As an evolution of the 
original method we propose an extension towards NN multiclass classification.

As a likelihood model, \DataModel comprises several hundreds of partially correlated
sources of systematic uncertainty for several hundreds of individual underlying 
rate measurements. For practical reasons, we have restricted ourselves to the \etau 
final state of the \PgtPgt system, with an electron (\Pe) and a hadronic \Pgt lepton 
decay (\tauh) in the final state and the data taken in 2017. For this subset of 
the data we use the same model as the one used in Ref.~\cite{CMS:2022kdi}. We 
compare both an inclusive and a simple differential STXS measurement based on \CENNT 
with \SANNT-based measurements taking 86 sources of partially correlated uncertainties 
into account under the most representative data-taking conditions for the year 2017.

The paper is organized as follows: Section~\ref{sec:experimental-setup} describes 
the experimental setup of the presented studies, including a brief description of 
the measurement used for benchmarking. Section~\ref{sec:sannt} describes the 
\SANNT, including a discussion of the concept, related work, and adaptations made 
for the studies presented here. An instructive comparison of the \SANNT with the 
original \CENNT on a simplified NN model for binary classification is given in 
Section~\ref{sec:comparison-cennt-sannt}. A conceptual extension of the loss 
function for multiclass classification is introduced in Section~\ref{sec:multiclass}, 
before we summarize the results in Section~\ref{sec:summary}.

\section{Experimental setup}\label{sec:experimental-setup}

\subsection{The CMS detector}\label{sec:CMS-detector}

The central feature of the CMS apparatus is a superconducting solenoid of 6\unit{m}
internal diameter, providing a magnetic field of 3.8\unit{T}. Within the solenoid 
volume are a silicon pixel and strip tracker, a lead tungstate crystal 
electromagnetic calorimeter, and a brass and scintillator hadron calorimeter, 
each composed of a barrel and two endcap sections. Forward calorimeters extend 
the pseudorapidity ($\eta$) coverage provided by the barrel and endcap detectors. 
Muons are detected in gas-ionization chambers embedded in the steel flux-return 
yoke outside the solenoid.

Events of interest are selected using a two-tiered trigger system. The first 
level, composed of custom hardware processors, uses information from the 
calorimeters and muon detectors to select events at a rate of around 100\unit{kHz} 
within a fixed latency of about 4\mus~\cite{Sirunyan:2020zal}. The second level, 
known as the high-level trigger, consists of a farm of processors running a 
version of the full event reconstruction software optimized for fast processing, 
and reduces the event rate to around 1\unit{kHz} before data 
storage~\cite{Khachatryan:2016bia,CMS:2024aqx}. 

A more detailed description of the CMS detector, together with a definition of 
the coordinate system used and the relevant kinematic variables, can be found 
in Ref.~\cite{Chatrchyan:2008zzk}. 

\subsection{Event reconstruction and selection}
\label{sec:event-reconstruction-and-selection}

The reconstruction of the proton-proton (\pp) collision products is based on the 
particle-flow (PF) algorithm~\cite{Sirunyan:2017ulk}, which combines the information 
from all CMS subdetectors to reconstruct a set of particle candidates, 
identified as charged hadrons, neutral hadrons, electrons, photons, and muons. The 
fully recorded detector data of a hard \pp collision defines an event for further 
processing. The primary vertex (PV) of the interaction is taken to be the vertex 
corresponding to the hardest scattering in an event, evaluated using tracking 
information alone, as described in Section 9.4.1 of Ref.~\cite{CMS-TDR-15-02}. 
Secondary vertices, which are detached from the PV, might be associated with decays 
of long-lived particles emerging from the PV. Any other collision vertices in an 
event are associated with additional, mostly soft inelastic \pp collisions called 
pileup.

{\tolerance=800
Electron candidates are reconstructed by combining clusters of energy deposits 
in the electromagnetic calorimeter with hits in the tracker~\cite{Khachatryan:2015hwa,
CMS:2020uim}. To increase their purity, reconstructed electrons are required 
to pass a multivariate electron identification discriminant, which combines 
information on track quality, shower shape, and kinematic quantities. Muons in 
the event are reconstructed performing a simultaneous track fit to hits in the 
tracker and in the muon chambers~\cite{CMS:2012nsv,CMS:2018rym}. For further 
characterization of an event, all reconstructed PF candidates are clustered into 
jets using the anti-\kt jet clustering algorithm, as implemented in the \FASTJET 
software package~\cite{Cacciari:2008gp,Cacciari:2011ma}, with a distance parameter 
of 0.4. Jets resulting from the hadronization of \PQb quarks are identified by 
exploiting dedicated identification algorithms, as described in Refs.~\cite{
Sirunyan:2017ezt,Bols:2020bkb}. Jets are also used as seeds for the identification 
of \tauh candidates. This is done by exploiting the substructure of the jets, 
using the ``hadrons-plus-strips'' algorithm, as described in Refs.~\cite{Sirunyan:2018pgf,
CMS:2022prd}. Decays into one or three charged hadrons with up to two neutral 
pions are used. To distinguish \tauh candidates from jets originating from the 
hadronization of quarks or gluons, as well as from electrons, or muons, the 
\textsc{DeepTau} algorithm~\cite{CMS:2022prd} is used, which provides three 
corresponding discriminants $D_{\alpha}\,(\alpha=\Pe,\,\PGm,\,\text{jet})$. 
For $D_{\Pe}$ the Tight, for $D_{\Pgm}$ the VLoose, and for $D_{\mathrm{jet}}$ 
the Tight working points, as defined in Ref.~\cite{CMS:2022prd}, are chosen. 
Eventually, the missing transverse momentum vector \ptvecmiss is calculated from 
the negative vector \pt sum of all PF candidates~\cite{Khachatryan:2014gga} 
exploiting the pileup-per-particle identification algorithm, as described in 
Ref.~\cite{Bertolini:2014bba}.
\par}

In the high-level trigger, events are selected either through the presence of both 
an electron and a \tauh candidate (\etau-trigger), or through the presence of 
only a high-\pt electron candidate (single-\Pe trigger). In the offline selection, 
the \Pe candidate is required to have $\abs{\eta}<2.1$ and $\pt>25\,(28)\GeV$, 
depending on whether the event was selected through the \etau-trigger (single-\Pe 
trigger) where the offline \pt threshold is chosen to ensure high efficiency 
of the trigger for the selected events. Furthermore, the \Pe candidate is required 
to be isolated from any hadronic activity originating from the PV. The \tauh 
candidate is required to have $\abs{\eta}<2.3$ and $\pt>35\,(30)\GeV$, depending 
on whether the event was selected through the \etau-trigger (single-\Pe trigger). 
The \Pe and \tauh candidates are required to be of opposite charges and separated 
by more than $\DR = 0.5$, where
  \begin{equation}
    \DR = \sqrt{\left(\Delta\eta\right)^{2}+\left(\Delta\phi\right)^{2}}
  \end{equation}
refers to the difference between two objects in the plane defined by $\eta$ and 
the azimuthal angle $\phi$, measured in radians. The presence of \Pe or \Pgm 
candidates in addition to the selected \Pe candidate is vetoed. A more detailed 
description of the selection criteria can be found in Ref.~\cite{CMS:2022kdi}.

\subsection{Data model}\label{sec:data-model}

After the selection described in Section~\ref{sec:event-reconstruction-and-selection}, 
referred to as the signal region (SR), the main backgrounds originate from \PZ 
boson production, also referred to as Drell--Yan (DY) production~\cite{PhysRevLett.25.316}, 
in association with jets in the \PgtPgt decay channel (\ZTT); \PW boson production 
in association with jets (\Wjets); top (\PQt) quark pair production (\ttbar); and 
SM events where light quark- or gluon-induced jets are produced through the strong 
interaction. This last background category is referred to here as quantum 
chromodynamics (QCD) multijet production. Minor backgrounds originate from the 
production of two vector bosons (diboson), single \PQt~quark, and \PZ boson 
production and decay into the \PePe and \PgmPgm final states (collectively denoted 
as \ZLL). Depending on the experimental signatures that these backgrounds leave 
in the detector, they are estimated either by the \Pgt-embedding method~\cite{
Sirunyan:2019drn}, the (``fake factor'') \FF-method~\cite{Sirunyan:2018qio,Sirunyan:2018zut}, 
or simulation to form the data model \DataModel to be used for signal extraction. 

In the \Pgt-embedding method, events with two muons are selected in data. All 
energy deposits of the muons are removed from the event record and replaced by 
the decay products and subsequent energy deposits of simulated \Pgt decays with 
the same kinematic properties as the selected muons. In this way, the method 
relies only on the simulation of the \Pgt decay and its energy deposits in the 
detector, while all other parts of an event are obtained from data. The resulting 
sample is used to estimate the background from all processes that comprise two 
genuine \Pgt leptons in their final states, such as \ZTT, corresponding decay 
channels of \ttbar, and diboson production. 

For the \FF-method, the selection criterion on $D_{\text{jet}}$ is inverted, 
keeping all other selection criteria unchanged, to obtain a sideband close to the 
SR. A minimal requirement on $D_{\text{jet}}$ is imposed to ensure kinematic  
proximity. Transfer functions \FF are applied to the selection to account for the
differences in the $D_{\text{jet}}$ selection. These \FF are obtained from 
independent determination regions in data. The \FF-method is used to estimate 
mostly the contributions of QCD multijet, \Wjets, and \ttbar production, where a 
quark- or gluon-induced jet is misidentified and selected as the \tauh candidate. 

{\tolerance=800
More than 80\% of all expected background processes are estimated either from the 
\Pgt-embedding method or the \FF-method and so are, at least partially, derived
from data. Remaining backgrounds and signal processes are estimated using simulated 
events generated by \mbox{\POWHEG 2}~\cite{Nason:2004rx,Frixione:2007vw,Alioli:2008tz,
Alioli:2010xd,Alioli:2010xa,Bagnaschi:2011tu} and \MGvATNLO 2.2.2 (2.4.2)~\cite{
Alwall:2011uj,MadGraph} in configurations described in Ref.~\cite{CMS:2022kdi}. 
The individual processes are normalized to their theoretical predictions with the 
highest available accuracy in the strong coupling constant \alpS. The event 
generator \PYTHIA~8.230~\cite{Sjostrand:2014zea} is used to simulate the parton
shower, as well as for the simulation of additional inclusive
inelastic \pp collisions, according to the expected pileup profile in data. All 
generated events are passed through a \GEANTfour-based~\cite{Agostinelli:2002hh} 
simulation of the CMS detector and reconstructed using the same version of the 
CMS event reconstruction software as used for the data. 
\par}

Control regions from event samples not included in the actual measurements are 
used to track how well the data can be described by \DataModel in terms of: 
object reconstruction; triggering, identification, and isolation efficiencies; 
object energy calibration scale factors; and general shape- and 
normalization-altering degrees of freedom of each individual background 
estimation method. Where needed, corrections are derived and applied as 
described in Ref.~\cite{CMS:2022kdi}. These corrections usually range below 10\%.

\subsection{Statistical inference of the signal}\label{sec:analysis-model}

Events in the SR are provided as input to a fully connected feed-forward NN for 
multiclass classification. In Ref.~\cite{CMS:2022kdi}, an NN with two hidden layers 
of 200 nodes and a hyperbolic tangent as the activation function for the hidden
nodes was used. For this paper, we have used a simpler configuration, with one 
hidden layer of 100 nodes and Rectified Linear Unit (ReLU)~\cite{Fukushima1975,   
10.5555/3104322.3104425} activation function, following the architecture that 
was chosen for the analysis of Ref.~\cite{Wunsch:2020iuh}. We favour comparability 
with respect to Ref.~\cite{Wunsch:2020iuh} over Ref.~\cite{CMS:2022kdi}, to allow 
for a direct comparison of the adaptations of the original method, as described in 
Section~\ref{sec:adaptations}.

The 14-dimensional vector of input features \vx to be passed to the NN for each 
event comprises: the \pt of both \Pgt candidates and their vector sum; the \pt 
of the two leading jets and their vector sum, their difference in $\eta$, and the 
dijet mass \mjj; the number of jets \Njet; the number of \PQb jets \Nbtag; a 
likelihood-based estimate of the mass of the \PgtPgt system \mtt~\cite{Bianchini:2016yrt}; 
the mass of the visible decay products of the \PgtPgt system \mvis; and estimates 
of the momentum transfer of each exchanged vector boson under the vector boson 
fusion (VBF) hypothesis, as used in Ref.~\cite{Gritsan:2016hjl}. 
The samples used for each corresponding NN training comprise events with fewer than 
two jets, for which input features like the \pt of the second leading jet or \mjj 
are not defined. In these cases, these observables are filled with predefined values 
outside their physical value spaces. While outside, the values are chosen still close 
to the physical value spaces, to minimize their influence on the NN prediction. 
Before entering the NN, all features are standardized such that their distributions 
have a mean of 0 and a standard deviation of 1.

The NN architecture has seven output nodes with NN output functions \yl, referring 
to the following event classes:
{\tolerance=800
\begin{itemize}
\item Events from background processes with two genuine $\Pgt$ decays in the final 
state, estimated from the $\Pgt$-embedding method (genuine $\Pgt$);
\item Events from processes where the $\tauh$ candidate originates from a 
misidentified quark- or gluon-induced jet, estimated from the \FF-method (\jettau
).
\item Events from \ttbar production, which are not covered by any of the above 
mentioned background estimation methods.
\item Events from \ZLL production.
\item Events from smaller background processes that are either difficult to 
isolate or too small to be treated individually for signal extraction. These  
are combined into one residual background class (misc).
\item Events from \PH production through gluon fusion (\ggH), as defined in the 
STXS scheme of the LHC Higgs Working Group~\cite{deFlorian:2016spz,Berger:2019wnu}.
\item Events from the the VBF dominated \qqH process, as defined in the STXS 
scheme of the LHC Higgs Working Group.
\end{itemize}
\par}

The output nodes correspond to the classes of the classification task. The 
activation function for the output nodes is the softmax function. The evaluation 
of the trained NN and signal extraction are performed on independent samples of 
\DataModel using a two-fold cross-validation scheme. 

For the signal extraction, each event is assigned to the class giving the highest 
value of \yl. Histogrammed distributions of \yl, for each corresponding class, 
are then used as input to an extended binned likelihood function of the form 
\ifthenelse{\boolean{cms@external}}
{
  \begin{multline}
      \LH\left(\{k_{i}\},\,\{\rs\},\,\{\thj\}\right) = \\
      \prod\limits_{i}\mathcal{P}\bigl(k_{i}|\lambda_{i}(\{\thj\})\bigr) 
      \prod\limits_{j}\mathcal{C}_{j}(\tilde{\theta}_{j}|\thj), 
      \label{eq:likelihood}
    \end{multline}

}
{
  \begin{equation}
      \LH\left(\{k_{i}\},\,\{\rs\},\,\{\thj\}\right) =
      \prod\limits_{i}\mathcal{P}\bigl(k_{i}|\lambda_{i}(\{\thj\})\bigr)\,
      \prod\limits_{j}\mathcal{C}_{j}(\tilde{\theta}_{j}|\thj),
    \label{eq:likelihood}
  \end{equation}
}
with 
  \begin{equation*}
      \lambda_{i}(\{\thj\}) = \sum\limits_{s}\rs\,S_{si}(\{\thj\})
      +\sum\limits_{b}B_{bi}(\{\thj\}),
  \end{equation*}
where $i$ labels all bins of the input distributions for each signal class (with 
index $s$) and background class (with index $b$) and the function \Poisson 
corresponds to the Poisson probability to observe $k_{i}$ events in bin $i$ for 
a prediction of $\sum\rs\,S_{si}$ signal and $\sum B_{bi}$ background events in 
that particular bin. The scaling parameters $\rs$ of the signal contributions 
$S_{si}$ relative to the SM expectation are the POIs. Systematic uncertainties 
affecting the predicted yields of $S_{si}$ and $B_{bi}$ in each bin are 
incorporated in the form of penalty terms for additional nuisance parameters 
\thj in the likelihood, entering as a product with predefined 
probability density functions $\mathcal{C}_{j}(\tilde{\theta}_{j}|\thj)$, 
where $\tilde{\theta}_{j}$ corresponds to the nominal value for \thj. 
Eventually, the signal is obtained from a maximum likelihood fit based on 
Eq.~(\ref{eq:likelihood}) to the data, during which the predefined uncertainties 
in the $\tilde{\theta}_{j}$ may be further constrained. This setup is identical 
to the one used in Ref.~\cite{CMS:2022kdi}. In the scope of this paper, the data 
$\{k_{i}\}$ are replaced by the Asimov data set $\{k_{i}^{\mathrm{A}}\}$ 
obtained from \DataModel including all $S_{s}$ and $B_{b}$ under the SM 
hypothesis with $\rs=1$, prior to any fit to the data. As introduced in 
Ref.~\cite{Cowan:2010js} the Asimov data set is the most representative outcome 
of a hypothetical experiment corresponding to the expectation value according to 
\DataModel in each possible observable.

For the discussion in Section~\ref{sec:comparison-cennt-sannt} we use a 
simplified setup, where all $S_{s}$ and all $B_{b}$ are combined 
into one single signal ($S$) and one single background ($B$) process each, 
resulting in a binary classification task to separate $S$ from $B$. In this case, 
a single NN output \y ranging from 0 (for $B$) to 1 (for $S$) is used and 
Eq.~(\ref{eq:likelihood}) is adapted accordingly. The final results presented in 
Sections~\ref{sec:comparison-cennt-sannt} and~\ref{sec:multiclass} are obtained 
using the CMS statistical analysis tool, \textsc{Combine}~\cite{CMS:2024onh}.

We note that $\hat{y}(\boldsymbol{\omega}, \vx)$ is a deterministic function of 
\vx and the NN weights $\boldsymbol{\omega}$. After training, the $\boldsymbol{
\omega}$ are fixed so that \vx is mapped to \y. We assume no additional 
uncertainty from the use of the explicit NN that has been trained for the given 
analysis task, such that all questions of uncertainties relate to the choice and 
quality of \DataModel and not to the NN itself. 

\subsection{Systematic uncertainties}\label{sec:uncertainty-model}

A detailed description of the original uncertainty model of \DataModel is given 
in Ref.~\cite{CMS:2022kdi}. The experimental uncertainties with the largest impact 
on the POI for an inclusive cross section measurement in the STXS setup based on 
the data set of the full Run 2 and all $\Pgt\Pgt$ final states are related to the 
$\Pgt$ identification efficiency $\etauID(40,500)$ for $40<\pt^{\tauh}<500\GeV$ 
and to the \FF-method. 

When restricted to the \etau final state and the data-taking year 2017, the 
uncertainty model of the STXS cross section measurement comprises 224 nuisance 
parameters, of which 127 are of experimental and 50 of theoretical origins. The 
former relate to uncertainties in object reconstruction, triggering, and 
identification, as well as in specific parts of the background model. The latter 
relate to process normalizations including uncertainties in the signal model, 
which because of the fact that they might apply only to parts of the kinematic 
phase space, may also have shape-altering effects. The remaining 47 nuisance 
parameters are related to the finite population of histogrammed template 
distributions available for signal extraction. 

In the original analysis, the last group of nuisance parameters has been 
incorporated for each bin of each corresponding template histogram, individually, 
following the approach proposed in Ref.~\cite{Barlow:1993dm}. Since they allow 
for independent variations of individual bins, they are also referred to as bin-by-bin 
uncertainties. All other uncertainties lead to correlated changes across bins 
taking the form of either normalization or bin-correlated, shape-altering variations. 
Depending on how they have been derived, correlations arise across individual 
signal and background samples, or individual uncertainties. The nuisance parameters 
most relevant for \SANNT belong to the following groups of uncertainties: 

\begin{itemize}
\item Uncertainties related to the reconstruction, triggering, and identification 
of the selected \Pe or \tauh candidates. 
\item Uncertainties related to the \FF-method.
\item Theoretical uncertainties related to the signal processes.
\item Individual uncertainties, such as an uncertainty in the reweighting of the 
simulated sample used for estimating the background from \ZLL production, in 
sideband regions of the data; an uncertainty in the misidentification rate of 
quark- or gluon-induced jets as electron candidates; and the uncertainty in the 
integrated luminosity~\cite{CMS-PAS-LUM-17-001,CMS-PAS-LUM-17-004,CMS-PAS-LUM-18-002,
CMS:2021xjt}. 
\end{itemize}

For practical reasons that are further described in Section~\ref{sec:methodology}, 
a total of 86 out of the original 224 nuisance parameters are considered for the 
statistical inference and \SANNT. These comprise shape-altering variations neglecting 
bin-by-bin uncertainties and a small number of important normalization changing 
uncertainties that can lead to shape altering effects, when combining individual 
processes. All other uncertainties are removed from \DataModel. This setup is 
compared to the benchmark using \CENNT with the same reduced uncertainty model 
and number of nuisance parameters.

\section{Systematic uncertainty aware neural network training}\label{sec:sannt}

\subsection{Emergence and methodology}\label{sec:methodology}

In this section we elaborate on the abstract work mechanism of a \CENNT-based 
measurement, as summarized in Section~\ref{sec:experimental-setup}, and where it 
falls short. Based on this discussion we discuss the emergence and key difference 
of \SANNT. Eventually, we point out the main challenges of \SANNT and how they have 
been addressed. Measurements discussed in the scope of this paper imply the 
estimation of one or more POIs of a statistical model, as defined in 
Eq.~(\ref{eq:likelihood}), from the comparison with data. Without loss of
generality, we restrict the discussion in this section to one POI that we call 
\rs. The measurement objective is to estimate \rs with minimal uncertainty \Drs 
and no bias, corresponding to the highest sensitivity of the measurement to this 
POI. 

\begin{figure*}[tbh]
  \centering
  \includegraphics[width=1.0\textwidth]{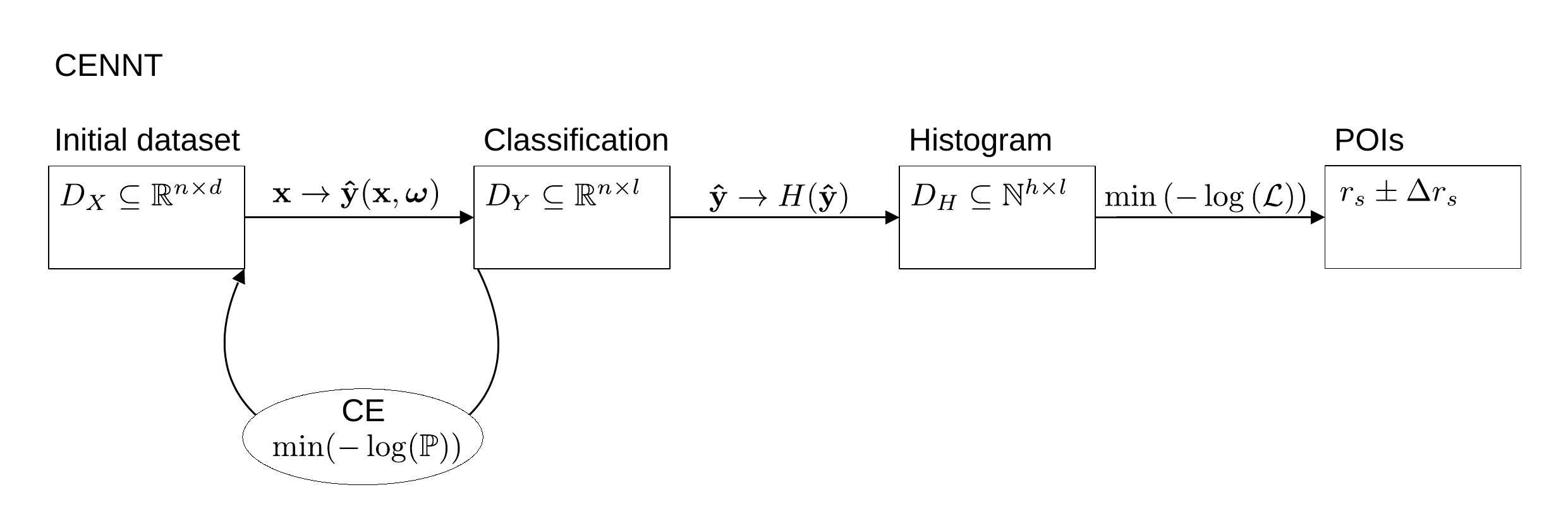}
  \includegraphics[width=1.0\textwidth]{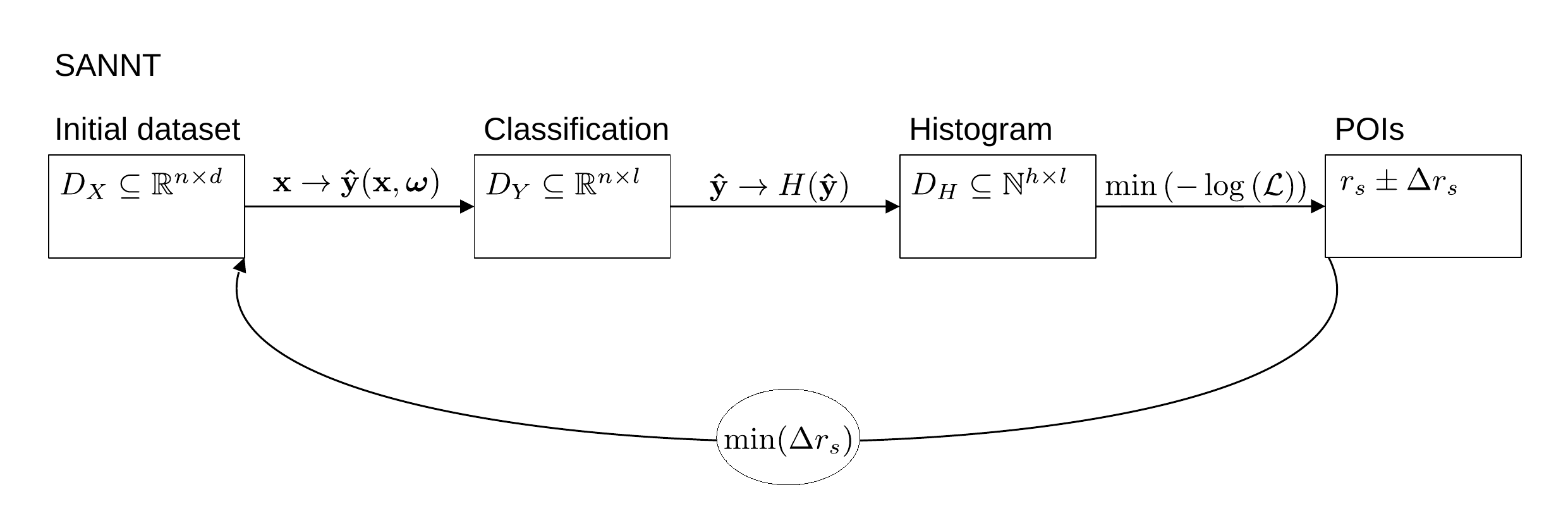}
  \caption{
    Flow chart of a \CENNT (upper) and \SANNT (lower). In the figure 
    $D_{i}$ denotes the data set; $n$ ($d$) the number of events (observables) in 
    the initial data set $D_{X}$; $l$ the number of classes after event classification; 
    and $h$ the number of histogram bins entering the statistical inference of 
    the POIs. The function symbol $\mathbb{P}$ represents the multinomial 
    distribution, and the symbol \LH has been defined in Eq.~(\ref{eq:likelihood}).
  }\label{fig:workflow}
\end{figure*}

The essence of statistical parameter estimation is to reduce a potentially large 
amount of information to a human-manageable and interpretable level. In the recent 
past, a sizable number of high-energy particle physics measurements have witnessed 
a significant increase in their sensitivity replacing traditional, human-inspired 
strategies for event classification by a \CENNT. A typical flow chart for such 
an analysis is illustrated in the upper part of Fig.~\ref{fig:workflow}. 
In a first step, the initial data set $D_{X}=\{\vx_0,\,\ldots,\,\vx_{n-1}\}\subseteq
\mathbb{R}^{d\times n}$ with $d$ characterizing observables $\vx\in X\subseteq 
\mathbb{R}^{d}$, is reduced to a lower-dimensional summary statistic $\mathbf{
\hat{y}}(\vx,\,\boldsymbol{\omega})\in Y\subseteq\mathbb{R}^{l}$, \eg through NN 
(multiclass) classification, where $l$ denotes the number of target classes and 
$\boldsymbol{\omega}$ refers to the trainable parameters of the NN model, to 
form a reduced data set $D_{Y}=\{\mathbf{\hat{y}}_{0}\left(\vx,\boldsymbol{\omega}
\right),\,\ldots,\,\mathbf{\hat{y}}_{l-1}\left(\vx,\boldsymbol{\omega}\right)\}
\subseteq \mathbb{R}^{l\times n}$. Since the \CENNT corresponds to the maximum 
likelihood estimate of \Multinomial, it is asymptotically efficient in fulfilling 
the separation task based on the provided $d$-dimensional feature 
space $X$~\cite{rao1992information,cramer1999mathematical}, so that the compression 
from $d$ to $l$ happens with minimal loss of information relevant for the 
separation. 

{\tolerance=800
For the analysis strategy discussed here, in a second step, the number of $n$ 
events is reduced to $h$ bins of a histogram 
\ifthenelse{\boolean{cms@external}}
{
  \begin{equation}
      H = 
      \left\{
        \left.\sum\limits_{j=0}^{n-1} H_{i}
        (\mathbf{\hat{y}}
          (\vx_{j},\boldsymbol{\omega}) 
        )
        \right| i=0\ldots h - 1
      \right\}, 
  \end{equation} 
}
{
  \begin{equation}
      H = 
      \left\{
        \left.\sum\limits_{j=0}^{n-1} H_{i}
        (\mathbf{\hat{y}}
          (\vx_{j},\boldsymbol{\omega}) 
        )
        \right| i=0\ldots h - 1
      \right\}, 
  \end{equation}
}
with
\begin{equation*}
  H_{i}\left(\mathbf{\hat{y}}\left(\vx,\boldsymbol{\omega}\right)
  \right) = \begin{cases}
  1, & \parbox[t]{.4\columnwidth}{if $\mathbf{\hat{y}}\left(\vx,\boldsymbol{
  \omega}\right)$ in the boundaries of bin $i$} \\
  0, &\text{otherwise},
  \end{cases}
\end{equation*}
resulting in an even further reduced data set $D_{H}=\{H_{0}(\mathbf{\hat{y}}
\left(\vx,\boldsymbol{\omega}\right)),\,\ldots,\,H_{h-1}(\mathbf{\hat{y}}\left(
\vx,\boldsymbol{\omega}\right))\}\subseteq \mathbb{N}^{l\times h}$. In this 
nomenclature we will refer to the Asimov data set $\{k_{i}^{\mathrm{A}}\}$ 
introduced in Section~\ref{sec:analysis-model} as \Asimov with $k_{i}^{
\mathcal{A}} = H_{i}(\mathbf{\hat{y}}\left(\vx^{\mathcal{A}},\boldsymbol{\omega
}\right))$, in the following.
\par}

A flaw of this strategy remains in the objective of the \CENNT, \ie, the separation 
of $S_{s}$ from $B_{b}$ based on \Multinomial, not being the same as the objective 
of the measurement, which is the minimization of \Drs. However, the ansatz remains 
successful, as long as the objectives of both estimates are approximately aligned, 
which can be assumed if the uncertainties in \rs are dominated by their statistical 
component \DrsStat and the yield of $S_{s}$ is small relative to $B_{b}$. 

The prime target of a \SANNT, as proposed, \eg, in Ref.~\cite{Wunsch:2020iuh}, is 
to become as congruent with the measurement objective as possible. From a given 
likelihood model, this can be achieved through the Fisher information 
matrix~\cite{fisher1922}
\ifthenelse{\boolean{cms@external}}
{
  \begin{equation}
    \label{eq:fisher_information_matrix}
    F_{\alpha_{i'}\alpha_{j'}} =
    \mathbb{E}
    \left[
      \frac{\partial^2}{\partial \alpha_{i'}\partial\alpha_{j'}}
      \Big( 
        -\log\LH
      \Big)
    \right]_{\hat{\alpha}_{i'}\,\hat{\alpha}_{j'}}, 
  \end{equation}
}
{
  \begin{equation}
    \label{eq:fisher_information_matrix}
    F_{\alpha_{i'}\alpha_{j'}} = \mathbb{E}
    \left[
      \frac{\partial^2}{\partial \alpha_{i'}\partial\alpha_{j'}}
      \Big( 
        -\log\LH
      \Big)
    \right]_{\hat{\alpha}_{i'}\,\hat{\alpha}_{j'}}, \\
  \end{equation}
}
with
\begin{equation*}
\alpha_{i'}\,\alpha_{j'}\in\{\{\rs\}, \{\thj\}\},\\
\end{equation*}
where in our case \LH is defined by Eq.~(\ref{eq:likelihood}) and $\hat{\alpha
}_{i}$ refers to the value of $\alpha_{i}$ that maximizes \LH. The expectation 
value is estimated from \Asimov, which is derived from the complete training 
sample, to minimize statistical fluctuations. Following the Cram\'{e}r-Rao 
inequality~\cite{cramer1999mathematical,rao1992information}, in the limit where 
the law of large numbers applies, \Drs can be obtained from 
  \begin{equation}
    \label{eq:variance_of_POIs}
    \Drs = \sqrt{F_{r_{s}r_{s}}^{-1}}.
  \end{equation}
This estimate of \Drs replaces CE as a loss function \LSANNT for \SANNT. The 
resulting modified flow chart is illustrated in the lower part of 
Fig.~\ref{fig:workflow}. 

The obvious advantage of this ansatz is that \LH may not only include the 
statistical, but also the systematic component of \Drs. Thus, it also leads to 
the most sensitive estimate based on $D_{H}$  for measurements that are not 
dominated by \DrsStat. On the downside, this ansatz has to address a number of 
challenges. It implies the computationally costly calculation and inversion of 
$F_{\alpha_{i}\alpha_{j}}$ during training. Moreover, it introduces an ambiguous 
choice of binning and implies the conceptual issue that \LH, as defined in 
Eq.~(\ref{eq:likelihood}), is not differentiable at its bin edges, a circumstance 
that prevents standard backpropagation through differentiation. One solution to 
this conceptional issue has been proposed in Ref.~\cite{inferno}, where the 
histogramming step is replaced by a modifiable softmax operator to approximate 
the histogram and sustain differentiability. In an ansatz followed in 
Ref.~\cite{Simpson:2022suz}, a kernel density estimate replaces the histogram. 
A likelihood ratio based on this approximation is computed and subsequently used 
as the objective for the NN training. 

At this point, we note that differentiability of \LSANNT is not strictly required
for backpropagation, as long as the information of proximity to a decision 
boundary, such as the bin edges of \LH in this case, is preserved. The solution 
proposed in Ref.~\cite{Wunsch:2020iuh} is to introduce a sum of custom functions 
\Bprop for the backward pass of the backpropagation instead of using the gradient. 
In the forward pass the loss function is evaluated as usual. The role of \Bprop is 
to allow movements of events across bin boundaries based on residual information 
about their positions inside a given bin. The exact form of \Bprop is a 
hyperparameter choice comparable to the choice of activation function for hidden 
nodes of the NN architecture. The same is true for the number of bins summarizing 
\yl to form the input to \LH, while we have not performed any particular 
optimization scan for this parameter.

The use of \LSANNT reveals a number of peculiarities, which are particularly
important for the studies presented in Sections~\ref{sec:comparison-cennt-sannt}
and~\ref{sec:multiclass}: 
\begin{itemize}
\item Considering $l$ as a signal, the choice of CE as loss function implies a 
strict ordering in \yl for events being background- or signal-like. The choice of 
\LSANNT does not imply such an ordering. Instead signal and background events 
obtain values of \yl that lead to the smallest value of \Drs. This can most easily 
be appreciated from Eq.~(\ref{eq:likelihood}), which is the basis for the calculation 
of \Drs. There, each bin is represented by a term $i$ under the first product. 
Since the scalar outputs of \Poisson commute under multiplication, the result of 
\LH does not depend on the order in which the bins are presented to \LH.   
\item A training with \Drs as objective is still fully supervised, but the event 
labels enter \LSANNT in a more subtle way than in the case of CE, viz.\ through 
the implicit modeling of $S_{s}$ and $B_{b}$. All processes are normalized according 
to their expected abundances in \Asimov.
\item A training with \Drs as objective only affects nuisance parameters \thj 
related to shape-altering variations of \yl. This can be best appreciated restricting 
the discussion to the optimization with respect to only one nuisance parameter. 
The parameters $\boldsymbol{\omega}$ forming the NN model induce variations of 
\yl. Thus, the concrete minimization of \Drs, which is tied to the values of 
$\boldsymbol{\omega}$, takes place through rearrangements of background and signal 
samples in the value space of \yl exploiting regions that are less affected by 
variations of \thj. If \thj has a pure normalization effect on \yl, the application 
of $\sum\Bprop$ has no effect on the minimization of \Drs, since the relative size 
of the variation is the same across the whole value space of \yl. We note at this 
point that normalization uncertainties may well be constrained during the 
maximum likelihood fit used for the extraction of the signal. 
\end{itemize}

We conclude this section with a note of caution. A \SANNT of any kind draws 
information from \DataModel including all the systematic variations in consideration.
The additional information from the uncertainty model leads to the reduction of 
\Drs compared to the outcome of a \CENNT. In turn, \SANNT implies stricter requirements 
on the quality and understanding of \DataModel. This can be appreciated by the 
following thoughts. We imagine a deficiency of \DataModel to describe the data 
in $X$, which is erroneously covered by a systematic variation that does not 
address the root cause of the deficiency. After \SANNT, \y might be derived from 
a region of $X$ where the erroneously covering variation is significantly smaller. 
As a consequence, \DataModel might reveal a discrepancy with data that can no longer 
be covered by the now reduced variation. Along these lines, we an\-ti\-ci\-pate 
that any random component in the estimation of a shape-altering systematic 
variation, \eg, caused by limited sizes of alternative samples used for estimating 
it, increases the variance of outcomes of a \SANNT, namely the reduction of \Drs. 
In the extreme of a purely random variation associated with \thrnd, events with 
larger or smaller associated variation would be randomly scattered in $X$. After 
\SANNT the variation associated with \thrnd would be minimized to best effort, 
while due to the random scattering of affected events in $X$ there is no reason 
why this minimization should have a positive influence on the reduction of \Drs. 
Rather, decreasing or increasing the importance of random subsets in $X$ for signal 
extraction should lead to a larger variance of potential outcomes of a \SANNT on 
\Drs. Both thoughts emphasize that special care should be taken to assure that 
the uncertainty model as an integral part of \DataModel be well defined and 
physically well motivated. While this is of obvious importance for any analysis, 
it is emphasized by the explicit use of the uncertainty model during a \SANNT.  

\subsection{Adaptations to the neural network training}\label{sec:adaptations}

In this section we discuss a small number of adaptations that we have made to 
the \SANNT, as originally proposed in Ref.~\cite{Wunsch:2020iuh}. For purposes 
of illustration, we base this discussion on the same set of pseudo-experiments as 
used in the original publication. This set consists of one signal and one 
background sample, each modeled by a multivariate Gaussian distribution centered 
around the points $(0,0)$ in the case of signal and $(1,1)$ in the case of 
background, in a two-dimensional feature space of a fully-connected feed-forward 
NN with features $x_{1}$ and $x_{2}$, one hidden layer of 100 nodes with ReLU 
activation function, and one single output $\y\in [0,1]$ for binary classification 
with the sigmoid activation function. Each Gaussian distribution is chosen with 
an identity covariance matrix. For the background sample, a systematic variation 
is introduced shifting the mean of the distribution by one unit along the 
direction of $x_{2}$.

\begin{figure*}[tbh]
  \centering
  \includegraphics[width=0.48\textwidth]{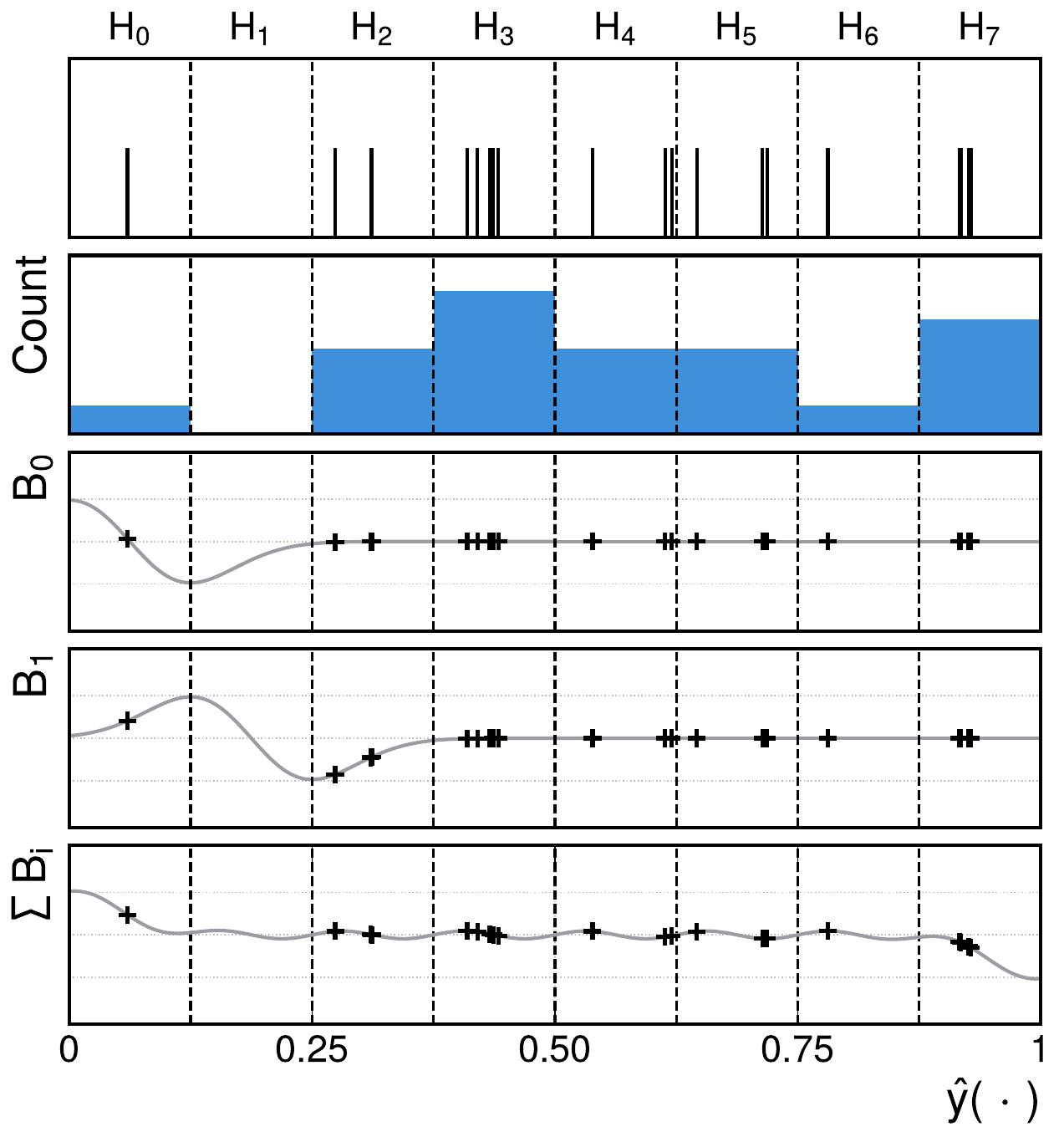}
  \includegraphics[width=0.48\textwidth]{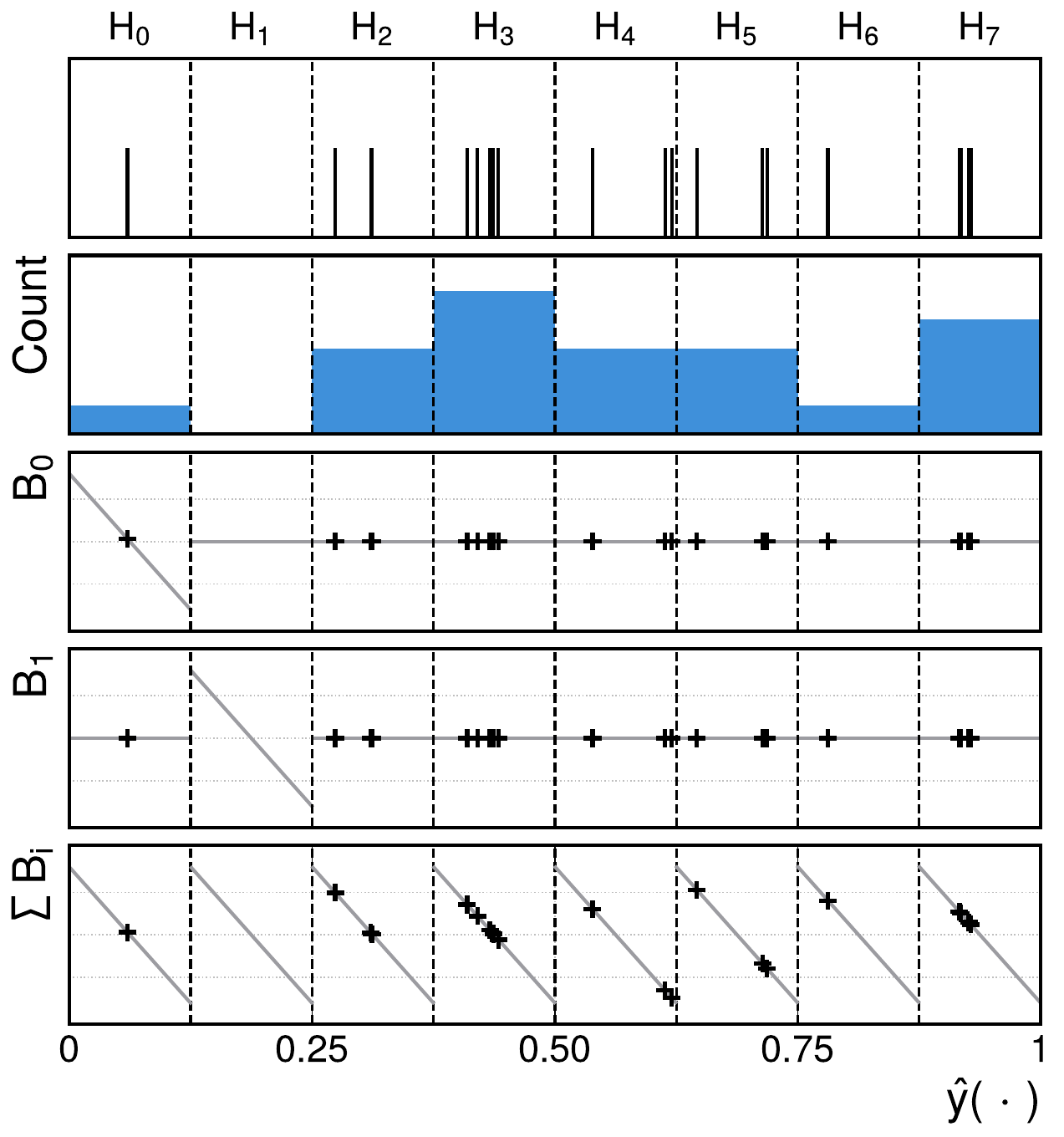}
  \caption{
    Custom functions \Bprop for the backward pass of the backpropagation algorithm, 
    as used (left) in Ref.~\cite{Wunsch:2020iuh} and (right) in this paper. In the 
    first row of each sub-figure, the same 20 random samples of a simple setup of 
    pseudo-experiments, as described in Section~\ref{sec:adaptations}, are shown. 
    In the second row the resulting histogram $H$, in the third and fourth rows 
    the functions $\mathcal{B}_{0}$ and $\mathcal{B}_{1}$ for the individual bins 
    $H_{0}$ and $H_{1}$, respectively, and in the last row the collective effect 
    of $\sum\Bprop$ are shown.
  }\label{fig:Bprop}
\end{figure*}

The original choice for \Bprop was the derivative of a Gaussian density with the 
expectation value $\mu_{i}$ at the center and standard deviation $\sigma_{i}$ 
corresponding to half the width of $H_{i}$. This choice drags samples away from 
the edges of $H_{i}$ leaving them untouched in the center. It is illustrated in 
Fig.~\ref{fig:Bprop} (left), where \y is summarized by a histogram with eight 
bins, the boundaries of which are indicated by the dashed vertical lines. 
The bars in the first row represent 20 randomly chosen samples of \y, in the 
second row, a snapshot of $H$, as obtained from these samples is shown. In a 
realistic setup, $H$ would be filled from a corresponding batch of the training 
sample. The following rows display $\mathcal{B}_{0}$ and $\mathcal{B}_{1}$ for 
$H_{0}$ and $H_{1}$, respectively. The last row shows the collective effect of 
$\sum\Bprop$ over all $H_{i}$. The intended effect per bin is clearly visible 
from the displays of $\mathcal{B}_{0}$ and $\mathcal{B}_{1}$. The display of 
$\sum\Bprop$ however also reveals undesired effects beyond the bins that the 
\Bprop are supposed to act upon, visible as interference effects with neighboring 
bins. In total, this leads to largely reduced values of $\sum\Bprop$ for the 
central bins of $H$ whereas the original magnitude of \Bprop is only retained at 
the lower and upper bounds of $H$. This further leads to reduced movements across 
bin edges with a tendency to drag events off the lower and upper bounding bins, 
towards the center of $H$. 

To make the backpropagation process more transparent and at the same time more 
comprehensible we have replaced the original choice of \Bprop by a simple linear 
function with intercept in the center of $H_{i}$. More importantly, we evaluate 
\Bprop only within the boundaries of $H_{i}$ and set it to zero otherwise. The 
effect of these modifications is shown in Fig.~\ref{fig:Bprop} (right). The 
desired effect per bin is preserved, movements across bin edges are better 
supported, and the lower and upper bounding bins are treated as equal to any other 
$H_{i}$. 

\begin{figure*}[tbh]
  \centering
  \includegraphics[width=0.48\textwidth]{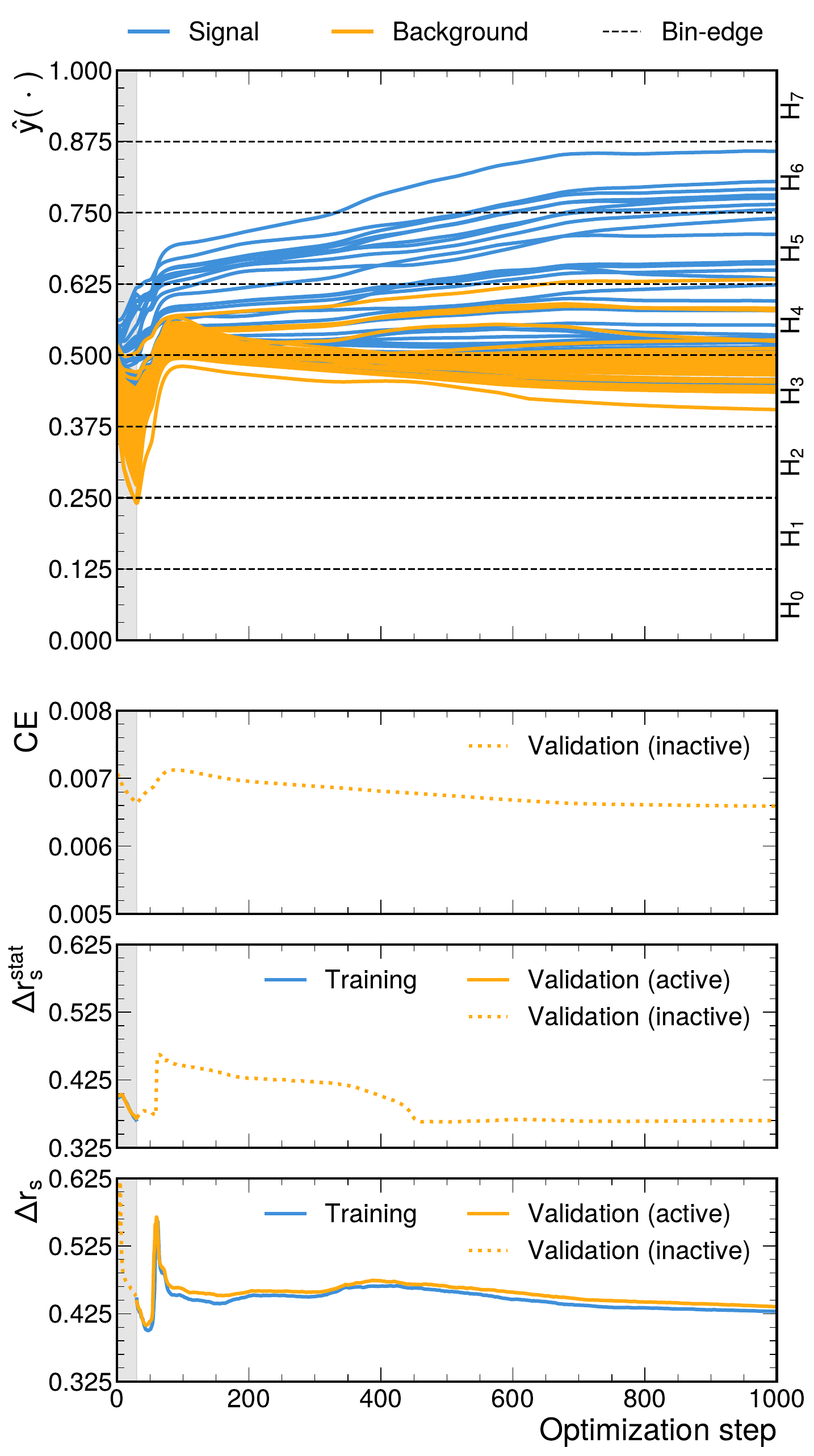}
  \includegraphics[width=0.48\textwidth]{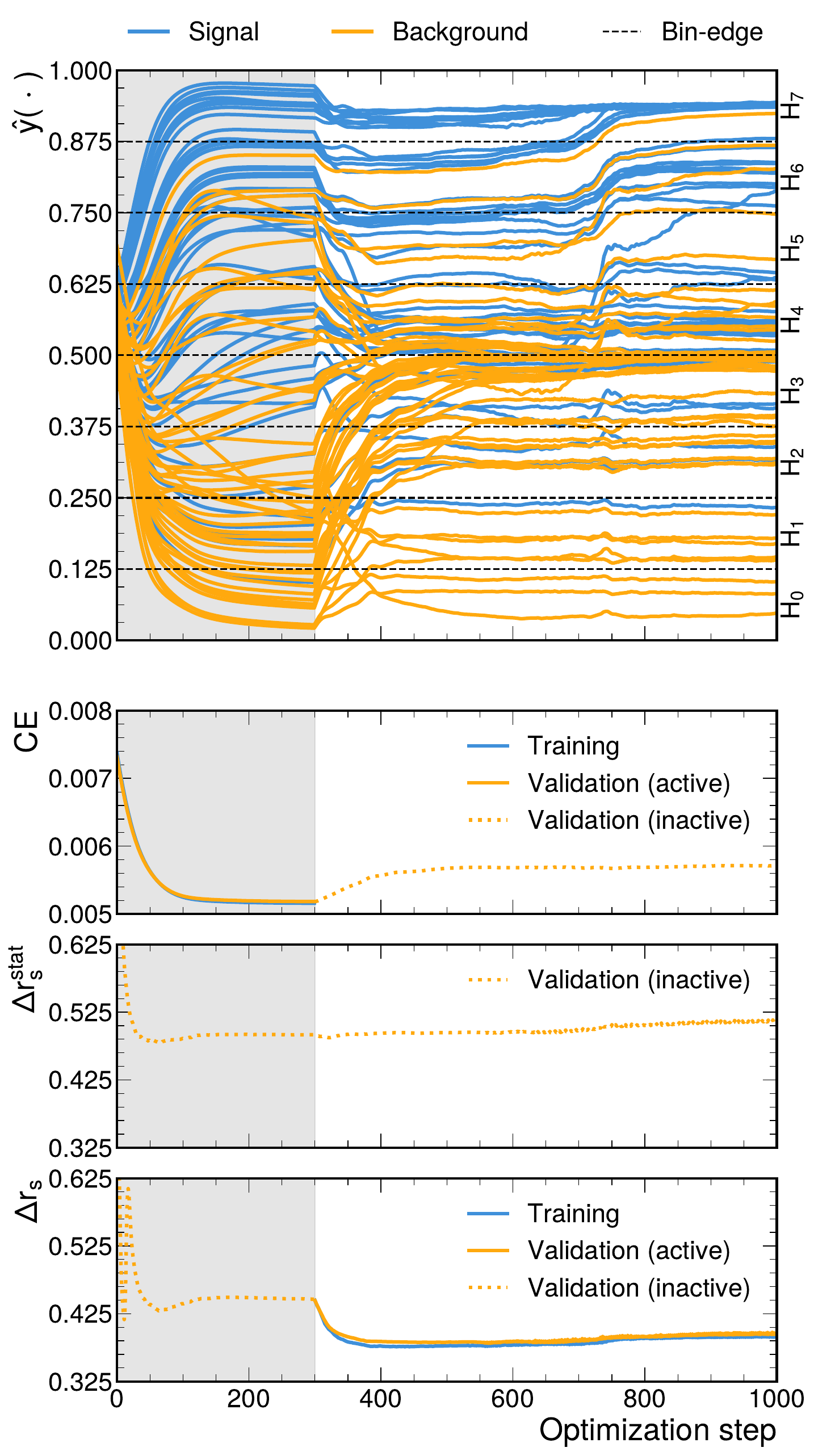}
  \caption{
    Evolution of the loss functions CE, \DrsStat, and \Drs as used (left) 
    in Ref.~\cite{Wunsch:2020iuh} and (right) for this paper. In the upper panels 
    the evolution of \y for randomly selected 50 (blue) signal and 50
    (orange) background samples during training is shown. Dashed horizontal lines 
    indicate the boundaries of the histogram bins $H_{i}$. The gray shaded area 
    indicates the pre-training. In the second and third panels from above the 
    evolution of CE and \DrsStat is shown. In the lowest panels the evolution 
    of $\LSANNT=\Drs$ is shown. The evaluation on the training (validation) 
    data set is indicated in blue (orange). The evaluation of the correspondingly 
    inactive loss function, during or after pre-training, evaluated on the 
    validation data set is indicated by the dashed orange curves. 
  }\label{fig:pretraining}
\end{figure*}
Another adaptation that we have made refers to the related initial observation 
in Ref.~\cite{Wunsch:2020iuh} that a training based only on \Drs tends to collapse 
\y into a usually very small number of central bins, after a few
optimization steps. To mitigate this effect and improve the training stability, 
a pre-training of 30 optimization steps, evaluating only \DrsStat as loss 
function, had been applied, prior to the full use of \LSANNT, in the original 
publication. 

We have replaced the loss function for pre-training by the binary CE, which all 
processes enter according to their expected abundance. In addition,we have 
extended the pre-training to 300 optimization steps to guarantee convergence. As 
discussed in Section~\ref{sec:methodology}, this modification should have an 
effect similar to the use only of \DrsStat, as shown in the third panel from above 
of Fig.~\ref{fig:pretraining} (right), where a reduction of \DrsStat during the 
CE-based pre-training is clearly visible. In addition, it introduces a separation 
of signal from background as an explicit pre-training objective and sustains a weak 
ordering of events being background- or signal-like equivalent to the CE-based 
pre-training. From the practical point of view, it leads to a wider spread of the 
samples across the available value space of \y. This strategy relies on the 
assumption that the result of a \CENNT, which coincides with the minimization only 
of \DrsStat, naturally forms a good starting point for the \SANNT. In the transition 
from a statistically to a systematically dominated measurement, the \SANNT acts as a 
continuous extension of the \CENNT. 

The evolution of \DrsStat and \Drs for a training based on the original setup 
and over 1000 optimization steps is shown in Fig.~\ref{fig:pretraining} (left). 
In the upper panel of the figure, the evolution of \y for randomly selected 50
(blue) signal and 50 (orange) background samples during training is shown, which 
we assume to be representative of the full background and signal samples. The 
optimization steps of each training are shown on the $x$ axis and the values of 
\y on the $y$ axis of the figure. The horizontal dashed lines indicate the bin edges 
of $H$, the gray shaded area indicates the pre-training. In Fig.~\ref{fig:pretraining} 
(right), the same evolution is shown for the modified setup. A much more pronounced 
spread of the samples across the value space of \y after pre-training is observed, 
which is sustained despite significant rearrangements, particularly of some 
signal samples, across histogram bins. For both setups, the value of \y stabilizes 
for each sample after roughly 500 optimization steps. 

In the lower part of the figure, the evolution of the loss functions over the same 
period of optimization steps is shown, split by (second panel) CE, (third panel) 
\DrsStat, and (last panel) $\LSANNT=\Drs$ using (left) \DrsStat and (right) CE 
during pre-training for \SANNT. The evaluations on the training and validation 
samples are indicated by the blue and orange curves, respectively. Beyond the last 
optimization step of the pre-training, from which on \DrsStat (CE) turns inactive, 
the blue line is not shown. It is still instructive to follow the evolution of 
\DrsStat and CE on the validation sample as indicated by the orange dashed lines. 
The complementary case for \Drs during pre-training and \SANNT is shown in the 
lower panels of the figure.

As \DrsStat and CE are based on different likelihood models their magnitudes
can not be directly compared. For \Drs, however, a direct comparison of the 
evolution by magnitude is valid. In both setups, the pre-training coincides
with a drop in \Drs. Convergence of the pre-training is more obviously reached 
in the case of CE. In both cases, the decrease in \Drs continues after pre-training, 
as intended. In the original setup however, a rearrangement of events across the 
very few bins, shown in the upper left part of the figure, leads to a significant 
increase in \DrsStat, visible in the third left panel from above. Shortly after, 
the increase in \DrsStat is outweighed by the decrease of the systematic component 
of \Drs. After 500 more optimization steps the model succeeds in slowly re-adapting 
to a minimal value of \DrsStat, similar to the one obtained after pre-training. A 
qualitatively similar, but more comprehensible progression is visible for the 
modified setup. The value of \Drs very quickly drops to a stable level within 100 
optimization steps. The decrease in the systematic contribution of \Drs happens 
at the cost of a slight increase in CE and \DrsStat. The increase of the latter
is less pronounced, which can be explained by the looser requirements on \DrsStat 
compared to CE, as discussed in Section~\ref{sec:methodology}.

The \Bprop might be viewed as an extension of the so-called straight-through 
estimator (STE) where \Bprop is replaced by the identity operation~\cite{STE}. We 
note that the choice of \Bprop features a number of important conceptual differences 
compared to the STE. Firstly, \Bprop acts only within the boundaries of bin $i$.
Secondly, and related to the first point, \Bprop supports explicit migrations 
across the bins of \LH, depending on the exact value of \y within the given bin. 
We identify these migrations as an essential ingredient to the successful 
minimization of \Drs. The evolution of a training equivalent to the one discussed 
here, with the only difference of replacing the \Bprop by identity operations is 
shown in Fig.~\ref{fig:ste-training} of Appendix~\ref{sec:appendix}. 
  
\section{Comparison with the cross-entropy-based training setup}
\label{sec:comparison-cennt-sannt}

It is instructive to compare the two training setups for a binary
classification task separating $S$ from $B$. For this purpose, all $S_{s}$ and all
$B_{b}$ (as discussed in Section~\ref{sec:data-model}) have been grouped into one 
sample each. 
\begin{figure*}[tbh]
  \centering
  \includegraphics[width=0.48\textwidth]{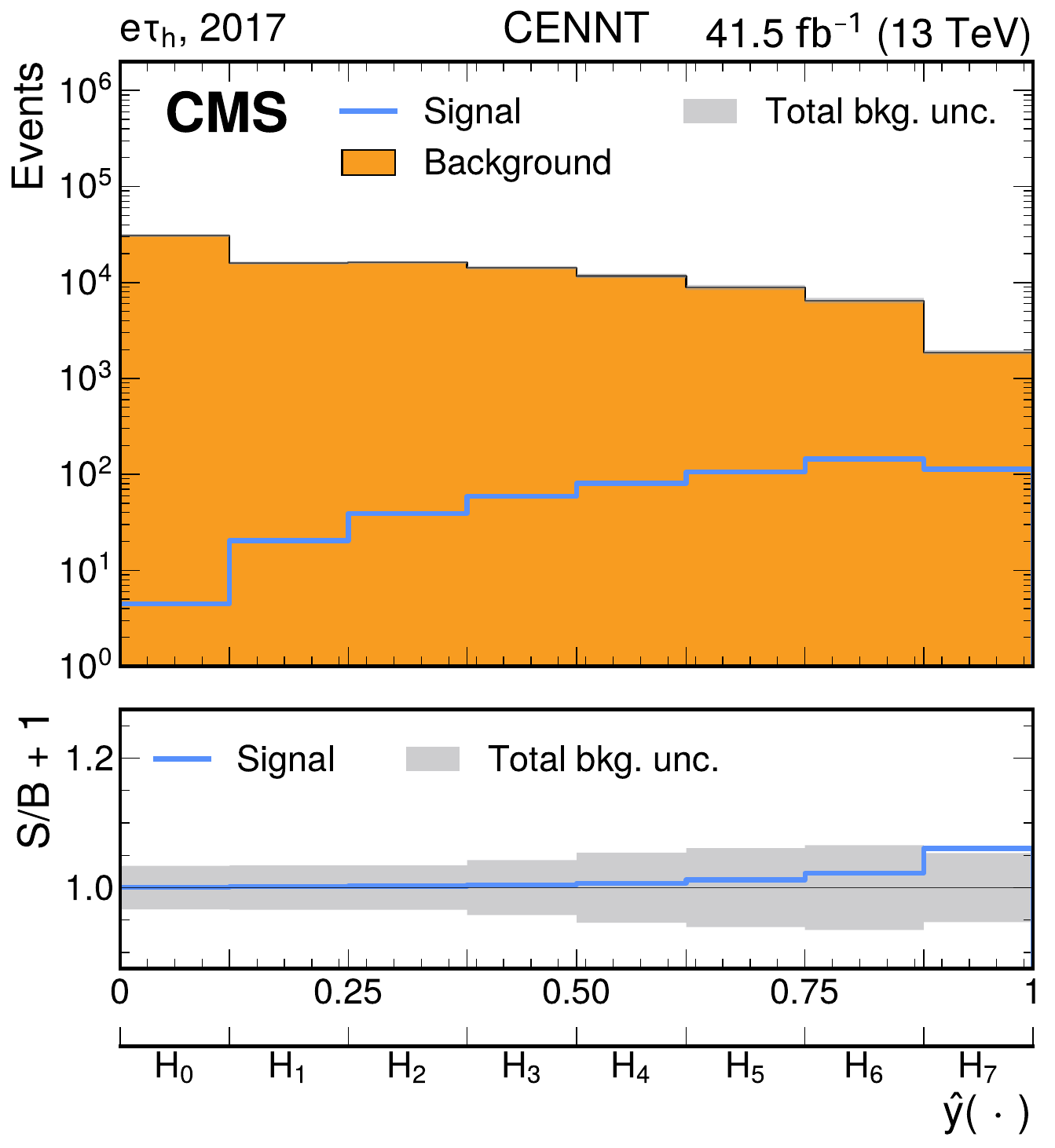}
  \includegraphics[width=0.48\textwidth]{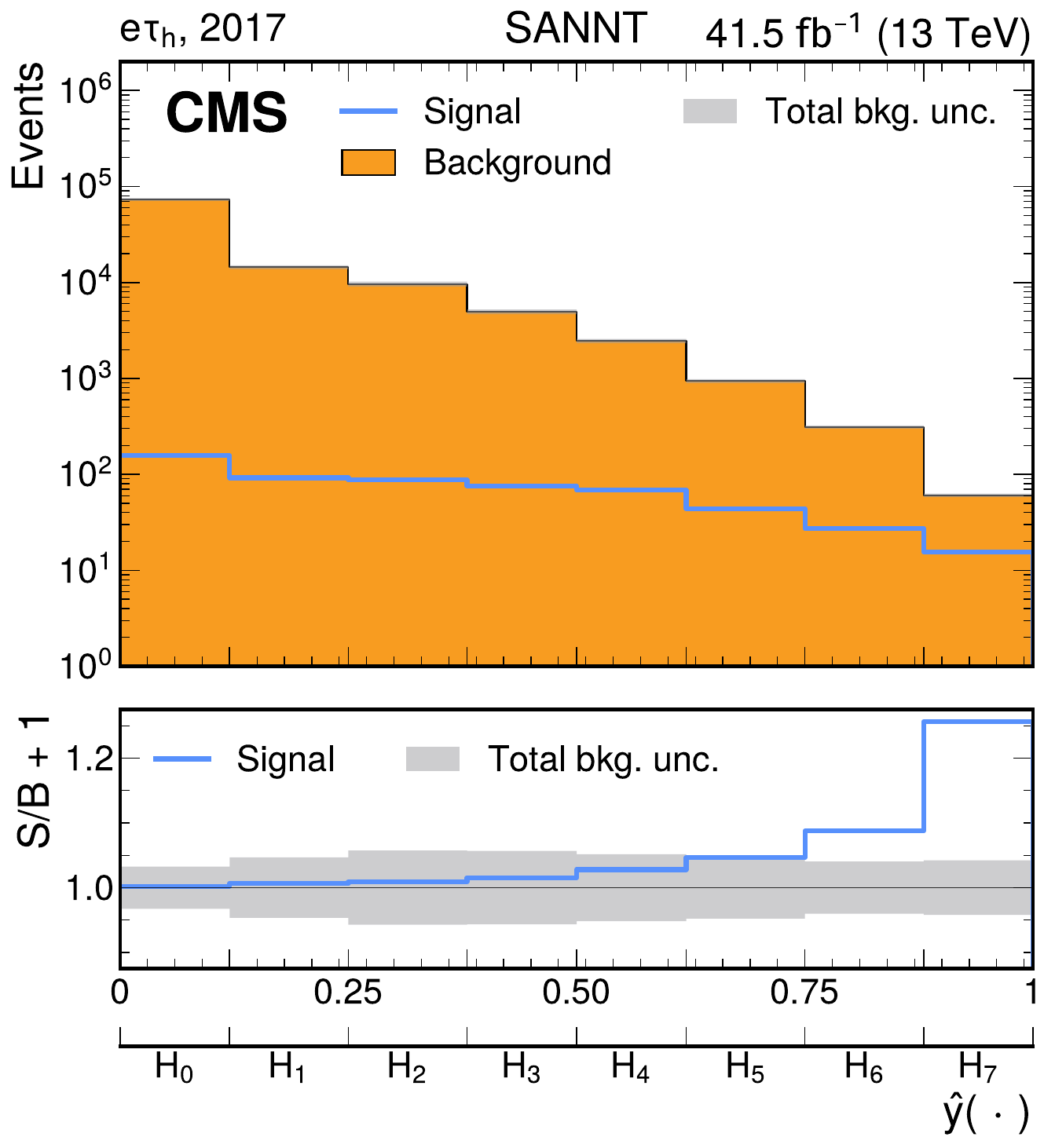}
  \caption{
    Expected distributions of \y for a binary classification task separating 
    $S$ from $B$, for a (left) \CENNT and (right) \SANNT, prior to any fit 
    to \Asimov. The individual distributions for $S$ and $B$ are shown by the 
    nonstacked open blue and filled orange histograms, respectively. In the 
    lower panels of the figures the expected values of $S/B+1$ are shown. The 
    gray bands correspond to the combined statistical and systematic uncertainty 
    in $B$. The boundaries of the histogram bins $H_{i}$ are also given on an 
    extra horizontal axis.
  }\label{fig:binary-discriminants}
\end{figure*}
The expected distributions of \y after \CENNT, prior to any fit to \Asimov, are 
shown in the upper panel of Fig.~\ref{fig:binary-discriminants} (left). The 
expectations for $S$ and $B$ are shown by the nonstacked open blue and the filled 
orange histogram, respectively. The expected value of $S/B+1$ is shown in the 
lower panel of the figure, where the gray band corresponds to the combined 
statistical and systematic uncertainty in $B$. A clear increase of $S/B$ is 
visible with the largest value in $H_{7}$ most to the right of $H$, where $S$ is 
expected to contribute ${\approx5\%}$ to the overall yield. The largest relative 
effect of the uncertainty in $B$ is expected in $H_{6}$. 

Fig.~\ref{fig:binary-discriminants} (right) shows the same distributions after
\SANNT. Compared to the \CENNT, both $S$ and $B$ are biased towards the left of 
the histogram and the expected largest relative effect of the uncertainties in $B$ is 
shifted towards $H_{2}$ and $H_{3}$. We note that after \SANNT there is no longer 
a strict ordering of $B$ to the left and $S$ to the right of $H$, while due to 
the pre-training this ordering is still weakly preserved. We have checked that the 
linear correlation coefficients between the output of \y after \CENNT and \SANNT 
are 0.80 for $S$ and 0.70 for $B$. This reduction in correlation relates to the 
impact that systematic variations have on \Drs. We anticipate that the larger this 
impact is, the smaller the correlations will be. 

This observation gives the following heuristic insights into the working mechanism 
of the \SANNT implemented here. Depending on their position in $X$, events in $D_{
X}$ are subject to systematic variations. For a well-defined classification task, 
$S$ and $B$ predominantly populate distinct subspaces $\SubS,\,\SubB\subset X$ 
with some potential transition region of overlapping events. For the given 
example, illustrated by Fig.~\ref{fig:binary-discriminants}, \y provides a mapping 
from $X$ to $\hat{y}\in[0,1]$. A fixed value of $\hat{y}=\kappa$ corresponds to 
the hyperplane of a discriminating boundary \DBkappa in $X$. Without loss of 
generality, a systematic ``up''-variation might move an event lying on \DBkappa 
and belonging to $B$ closer to \SubS. The ``down''-variation in turn might move 
a similar event lying on \DBkappa and belonging to $S$ closer to \SubB. We observe 
that the elements of \vx are weighted such that these events will obtain values 
of $\hat{y}\lesssim\kappa$, after \SANNT. This explains the two observations that 
can be made from Fig.~\ref{fig:binary-discriminants}: i) the general shift of \y 
towards smaller values of $\hat{y}$ as well as; ii) the larger values of $S/B+1$ 
for the events in bins $H_{6}$ and $H_{7}$, visible from the lower panel of 
Fig.~\ref{fig:binary-discriminants} (right). There may well be more subtle
effects steering the exploitation of features \vx by the NN model playing a role 
into this observation.

To visualize the influence of single features and their pairwise linear correlations, 
obtained from the support of $S$ and $B$ during training, on \y we have used the 
metric
  \begin{equation}
    \label{eq:talpha}
    \talpha = \frac{1}{n}\sum\limits_{i=1}^{n}\abs{t_{\alpha}(\vx_{i})}
  \end{equation}
based on a Taylor expansion of \y, with respect to the elements of \vx up to 
second order, as described in Ref.~\cite{Wunsch:2018oxb}. In Eq.~(\ref{eq:talpha}), 
$\alpha$ refers to a power set of the components of \vx, where elements may also 
be repeated, $t_{\alpha}$ refers to the Taylor series coefficient for the derivative 
of \y with respect to $\alpha$, and $n$ to the size of the training sample. Large 
(small) values of \talpha indicate that feature $\alpha$ has a large (small) 
influence on \y. For the \SANNT we have replaced the derivative of \LH in the 
Taylor expansion by $\sum\Bprop$. 
\begin{figure*}[tbh]
  \centering
  \includegraphics[width=0.8\textwidth]{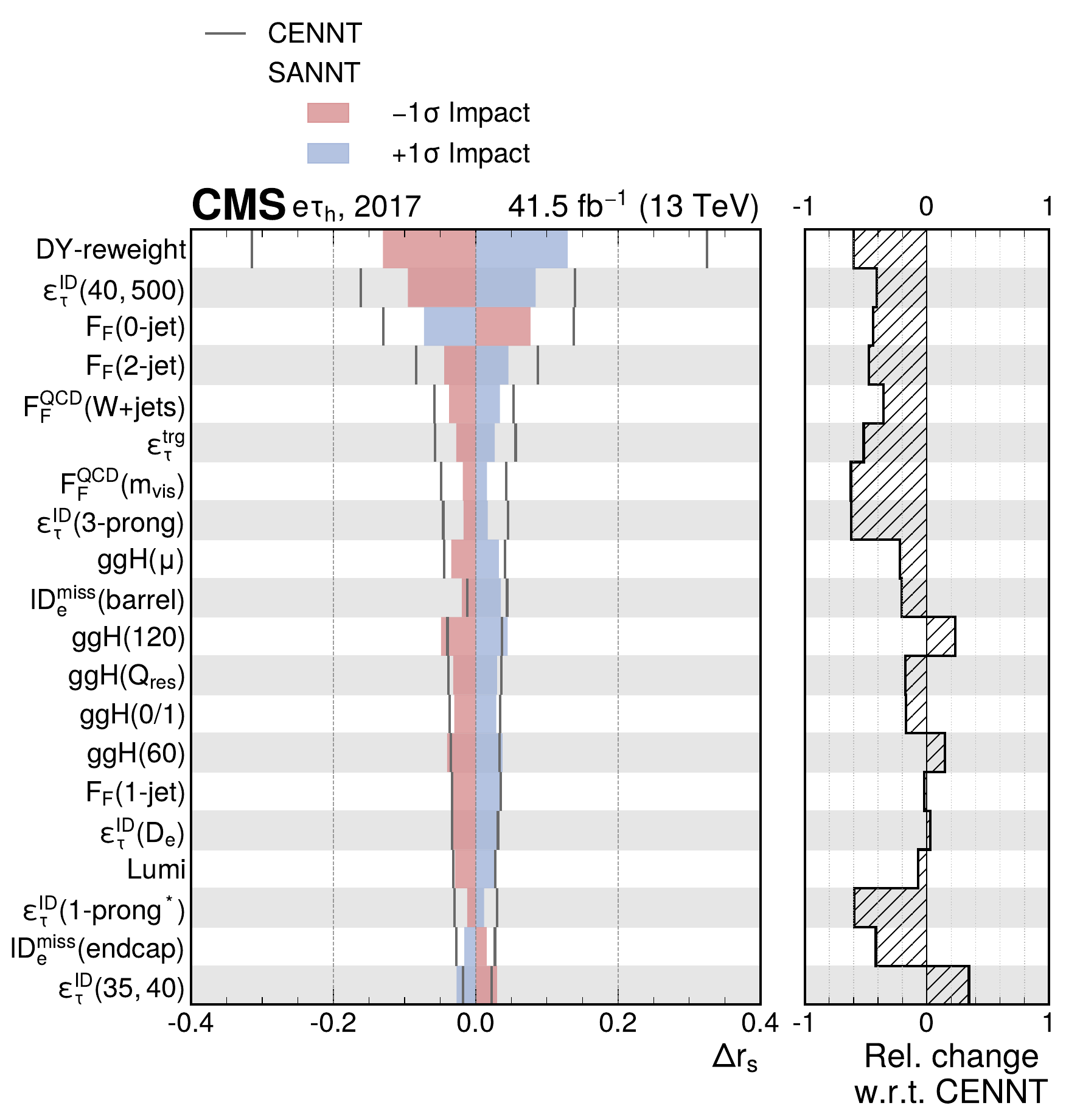}
  \caption{
    Impacts for the 20 nuisance parameters \thj with the largest impacts on \rs. 
    The gray lines refer to the \CENNT and the colored bars to the \SANNT.  The 
    impacts can be read from the $x$ axis. Labels shown on the $y$ axis for each 
    \thj are defined in Table~\ref{tab:binary-variations}. The entries are ordered 
    by decreasing magnitude for \CENNT when moving from the top to the bottom of 
    the figure. The panel on the right shows the relative change of the symmetrized 
    impact when moving from \CENNT to \SANNT. A more detailed discussion is given 
    in the text.
  }\label{fig:binary-NPs}
\end{figure*}

From this study, we have identified that especially the first-order coefficients 
for $\alpha=\mtt$ and \mvis and the second-order coefficients for $\alpha=(\mvis,
\mvis)$ and $(\mtt,\mtt)$ have a large influence on \y, after \CENNT. The 
coefficients $t_{\mtt}$ and $t_{\mvis}$ indicate the location in the marginal 
distributions of $S$ and $B$; the coefficients $t_{\mtt,\mtt}$ and $t_{\mvis,
\mvis}$ indicate the curvatures of each distribution and thus how much it is 
peaking or not. Despite the simplifications of the analysis, these findings, 
which follow our physics intuition, are in good accord with the findings of 
Ref.~\cite{CMS:2022kdi}. In the transition from \CENNT to \SANNT, we observe that 
the influences of the properties associated with $t_{\mvis,\mvis}$, $t_{\mvis}$, 
and $t_{\mtt}$ are largely reduced in favor of the curvature of \mtt, associated 
with $t_{\mtt,\mtt}$. 

The 20 nuisance parameters with the largest impacts on \rs, obtained from fitting 
the $S+B$-model with $\rs=1$ to \Asimov after shifting each parameter \thj 
individually by one standard deviation up and down fixing all other $\{\theta_{i
};\,i\neq j\}$ to their most probable values after fit, are shown in 
Fig.~\ref{fig:binary-NPs}. The impacts can be read off from the $x$ axis. The 
gray lines refer to the \CENNT and the colored bars to the \SANNT. Labels for 
each \thj, decreasing in magnitude for the \CENNT when moving from upper to lower 
entries in the figure, are given on the $y$ axis, referring to the following sets 
of systematic uncertainties:
{\tolerance=1200
\begin{itemize}
\item
Uncertainties mostly related to the \Pgt-embedding method, viz.\ in (\etauTrig) 
the efficiency to trigger the \etau pair; ($\etauID(D_{\Pe})$) the efficiency to 
distinguish \tauh candidates from electrons; and the efficiency to distinguish 
\tauh candidates with ($\etauID(35,40)$) $35<\pt^{\tauh}<40\GeV$ and ($\etauID(
40,500)$) $40<\pt^{\tauh}<500\GeV$ from quark- or gluon-induced jets. For 
$\etauID(40,500)$ additional uncorrelated parts for \tauh decays ($\etauID(
\text{3-prong})$) with three charged pions, and ($\etauID(\text{1-prong}^{*})$) 
one charged and additional neutral pions appear in the list.
\item 
Uncertainties related to the \FF-method, viz.\ in the normalizations of the 
estimates of events with zero, one, and two jets corresponding to 
$\FF(\text{0-jet})$, $\FF(\text{1-jet})$, and $\FF(\text{2-jet})$, respectively; 
the nonclosure correction to the estimate of the background contribution from QCD 
multijet production as a function of \mvis, $\FF^{\mathrm{QCD}}(\mvis)$; and the 
subtraction of \Wjets events as obtained from simulation during the estimation of 
the same background contribution, $\FF^{\mathrm{QCD}}(\Wjets)$.  
\item
Uncertainties related to simulation, viz.\ in the reweighting of the sample used 
for estimating the background from \ZLL production, in sideband regions of the 
data (DY reweight); the misidentification rate of quark- or gluon-induced jets 
as electron candidates in the ($\missIDe(\mathrm{barrel})$) barrel and ($\missIDe
(\mathrm{endcap})$) endcap regions of the detector; and the integrated luminosity 
(Lumi). 
\item
A number of uncertainties related to the modeling of the \ggH signal, viz.\ from 
variations of ($\ggH(\mu)$) the ($\mu_{\mathrm{r}}$) renormalization and ($\mu_{
\mathrm{f}}$) factorization, as well as ($\ggH(Q_{\mathrm{res}})$) the resummation 
scales; migration effects across STXS bins ($\ggH(0/1)$) with 0 and 1 jet, as 
well as the $\pt^{\PH}$ STXS bin boundaries at ($\ggH(60)$) 60 and ($\ggH(120)$) 
120\GeV. 
\end{itemize}
\par}

\begin{table*}[tbh]
  \centering
  \topcaption{
    Association of nuisance parameters \thj with the systematic 
    variations they refer to for the 20 parameters with the largest impacts 
    on \rs, as shown in Fig.~\ref{fig:binary-NPs}. The label of each corresponding 
    uncertainty is given in the first column, the type of uncertainty, process 
    that it applies to, and rank in Fig.~\ref{fig:binary-NPs} are given in the 
    second, third, and fourth columns, respectively. The ``*'' in $\etauID(
    \text{1-prong}^{*})$ refers to the fact that this is the decay channel with 
    neutral pions in addition to the charged prong. The symbol EMB refers to 
    \Pgt-embedded event samples. A more detailed discussion is given in the 
    text.
  }
  \begin{tabular}{llcrccl}
    Label & Type & Process & \multicolumn{1}{c}{Rank} & Norm & Shape & Comment \\
    \hline
    \etauTrig                     & $\PGt$ trigger  & EMB     & 6 & \NA  & $\checkmark$ & \NA                      \\ 
    $\etauID(D_{\Pe})$            & $\PGt$ ID       & MC, EMB & 16              & $\checkmark$  & \NA & Discr.\ against \Pe       \\
    $\etauID(35,40)$              & $\PGt$ ID       & EMB     & 20              & \NA  & $\checkmark$ & $35<\pt^{\tauh}<40\GeV$  \\
    $\etauID(40,500)$             & $\PGt$ ID       & EMB     & 2 & \NA  & $\checkmark$ & $40<\pt^{\tauh}<500\GeV$ \\
    $\etauID(\text{1-prong}^{*})$ & $\PGt$ ID       & EMB     & 18              & \NA  & $\checkmark$ & One $\pi^{+}$+$\pi^{0}$'s\\
    $\etauID(\text{3-prong})$     & $\PGt$ ID       & EMB     & 8 & \NA  & $\checkmark$ & Three $\pi^{+}$'s        \\ [\cmsTabSkip]
    $\FF(\text{0-jet})$           & Norm.           & \FF     & 3 & \NA  & $\checkmark$ & $\Njet=0$                \\
    $\FF(\text{1-jet})$           & Norm.           & \FF     & 15              & \NA  & $\checkmark$ & $\Njet=1$                \\
    $\FF(\text{2-jet})$           & Norm.           & \FF     & 4 & \NA  & $\checkmark$ & $\Njet=2$                \\
    $\FF^{\mathrm{QCD}}(\mvis)$   & Nonclosure      & \FF     & 7 & \NA  & $\checkmark$ & In \mvis                 \\
    $\FF^{\mathrm{QCD}}(\Wjets)$  & Subtr.\          & \FF     & 5 & \NA  & $\checkmark$ & Subtr.\ of sim.             \\ [\cmsTabSkip]
    $\ggH(\mu)$                   & Theory          & \ggH    & 9 & \NA  & $\checkmark$ & $\mu_{\mathrm{r}}$ and $\mu_{\mathrm{f}}$  \\
    $\ggH(Q_{\mathrm{res}})$      & Theory          & \ggH    & 12              & \NA  & $\checkmark$ & Resummation              \\
    $\ggH(0/1)$                   & Theory          & \ggH    & 13              & \NA  & $\checkmark$ & $0\to1$ jet migr.        \\
    $\ggH(60)$                    & Theory          & \ggH    & 14              & \NA  & $\checkmark$ & $\pt^{\PH}$ migr.        \\
    $\ggH(120)$                   & Theory          & \ggH    & 11              & \NA  & $\checkmark$ & $\pt^{\PH}$ migr.        \\ [\cmsTabSkip]
    $\missIDe(\mathrm{barrel})$   & \Pe misID       & MC      & 10              & \NA  & $\checkmark$ & Barrel                   \\
    $\missIDe(\mathrm{endcap})$   & \Pe misID       & MC      & 19              & \NA  & $\checkmark$ & Endcap                   \\
    DY reweight                   & Reweight        & MC      & 1 & \NA  & $\checkmark$ & In $\pt^{\Pgm\Pgm}$ and $m_{\Pgm\Pgm}$ \\
    Lumi                          & Int.\ luminosity & MC      & 17              & $\checkmark$  & \NA & \NA                      \\
    \hline 
  \end{tabular}\label{tab:binary-variations}
\end{table*}

{\tolerance=800
A summary of these variations is given in Table~\ref{tab:binary-variations}. The 
largest impact of a single uncertainty on \rs for \CENNT is observed for 
DY reweight, followed by $\etauID(40,500)$ and $\FF(\text{0-jet})$. With a few 
exceptions, all uncertainties are significantly reduced by the \SANNT demonstrating 
the success of the ansatz. The uncertainty due to DY reweight, with the largest 
impact on \rs, is reduced by nearly 60\%. The $\PZ\to\Pe\Pe$ process, related to 
this uncertainty, is closely located to $S$ in \mvis. The reduction of the DY 
reweight uncertainty is related to the reduced influence of the location and 
curvature of \mvis on \y previously discussed in this section. 
\par}

\begin{figure}[tbh]
  \centering
  \includegraphics[width=0.48\textwidth]{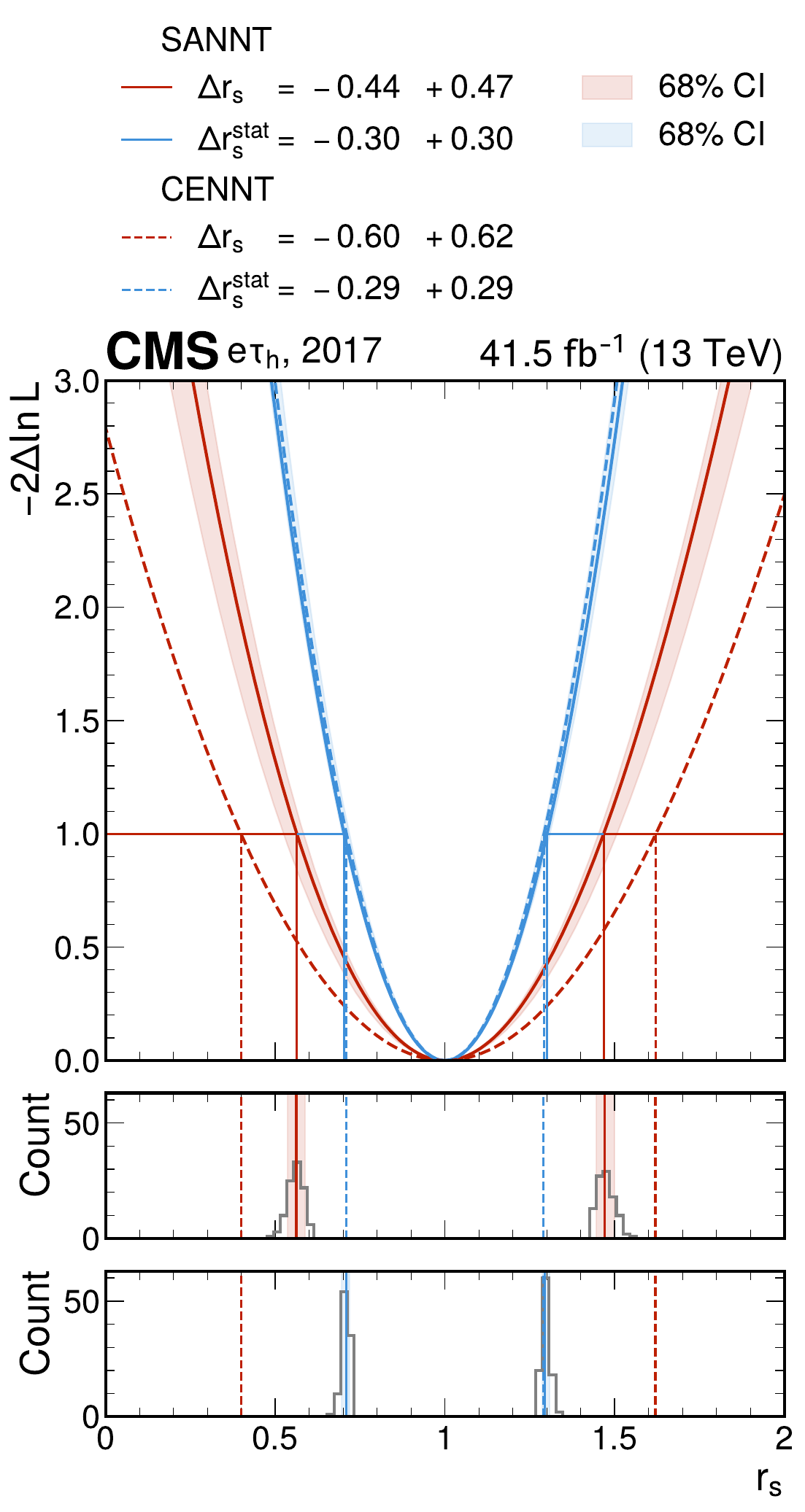}
  \caption{
    Negative log of the profile likelihood \NLL as a function of \rs, taking into 
    account (red) all and (blue) only the statistical uncertainties in \Drs. The 
    results obtained from the \CENNT are indicated by the dashed lines, and the 
    median expected results of ensembles of 100 repetitions of the \SANNT varying 
    random initializations are indicated by the continuous lines. The red and blue 
    shaded bands surrounding the median expectations indicate 68\% confidence 
    intervals (CI) from these ensembles. The lower panels show the distributions 
    underlying these CI.
  }\label{fig:binary-NLL-scan}
\end{figure}

Figure~\ref{fig:binary-NLL-scan} (upper) shows distributions of the negative 
log of the profile likelihood \NLL as a function of \rs, based on the input 
distributions as depicted in Fig.~\ref{fig:binary-discriminants}. For the red 
(blue) curves all (only the statistical) uncertainties in \rs have been taken 
into account.
For the latter, all nuisance parameters have been fixed to their most probable 
values after fit, before performing a second fit to \Asimov. 

The most probable value of \rs is observed at 1, as expected from a fit to \Asimov 
demonstrating that the estimate is unbiased. The values of \Drs and \DrsStat are 
obtained from the intervals in \rs with $\NLL<1$. The results as obtained from 
the \CENNT are indicated by the dashed lines. The continuous lines indicate the 
median results of ensembles of 100 independent repetitions each of the \SANNT 
varying random initializations. The red and blue shaded bands surrounding the 
median expectations indicate the 68\% central intervals of these ensembles, 
representing the variances of the obtained results due to random choices for 
training. In the lower panels, the underlying distributions to these central intervals 
are shown in red and blue, correspondingly. Based on \SANNT (\CENNT) values of 
$\Drs=^{+0.47}_{-0.44}$ ($^{+0.62}_{-0.60}$) are obtained, corresponding to a 
reduction of 25\% in magnitude for the result obtained from \SANNT with respect 
to the \CENNT. Taking only the statistical component of \Drs into account, a 
value of $\DrsStat=^{+0.31}_{-0.30}$ ($\pm0.29$) for \SANNT (\CENNT) is obtained. 
This indicates that the \SANNT is a significant improvement to the \CENNT. It is 
the first time that the success of a \SANNT has been demonstrated for an analysis 
at this level of complexity.

\section{Extension for multiclass classification}\label{sec:multiclass}

In this section we report the application of \SANNT to the multiclass-classification 
task, based on the original STXS cross section measurement described in 
Ref.~\cite{CMS:2022kdi} exploiting the five background and two signal classes, 
as described in Section~\ref{sec:analysis-model}. Each class $l$ is represented 
by an output node of the NN with output function \yl. The effect of multiclass 
classification is achieved through the use of a softmax operator in the output 
layer of the NN providing a probability estimate for an event with a feature 
vector \vx to originate from $l$. The \CENNT is obtained from a simple extension 
of binary to categorical CE. 

Naively the same replacement of binary by categorical CE would apply for the 
\SANNT pre-training. Instead, we keep the sigmoid activation function, applying 
it to all output nodes and exploit the custom function 
  \begin{equation}
    \mathrm{CE'} = -\sum\limits_{i}\left(w_{i}
    \sum\limits_{l} \mathrm{CE}_{l}\right),
  \end{equation}
where $\mathrm{CE}_{l}$ corresponds to the binary CE evaluated separately for 
each $l$, $i$ runs over all events in \Asimov, and $w_{i}$ normalizes each event 
to the expectation for the process it belongs to and the integrated luminosity 
of the assumed data set. We have made this modification as we observed a significant
improvement of the subsequent \SANNT, which as in the previous case of binary 
classification, we attribute to the practical effect of a wider spread of samples 
across the value spaces of \yl. 

As discussed at the end of Section~\ref{sec:methodology}, the fact that any binned 
likelihood function, \eg, as defined in Eq.~(\ref{eq:likelihood}), lacks immanent
information about the relations between bins, necessarily implies the loss of an 
ordering principle for the association of events of class $l$ with a given value of 
\yl. More importantly in the case of multiclass classification, this even prevents 
a fixed association of events to $l$. While this circumstance does not pose an 
issue of principle for a multiclass-classification \SANNT, it renders the conventional 
concept of output classes for multiclass classification meaningless, since, for 
example, there is no rule in the training to prevent events of class ``genuine \PGt'' 
from being associated with the output node meant to collect events of class ``misc'' 
and vice versa. Even worse, both event classes could be associated with the same 
output node.

In Section~\ref{sec:adaptations} we have introduced the result of a \CENNT 
pre-training as natural starting point for the \SANNT, preserving a weak relation 
between the class that an event belongs to and the value of \y, in the transition 
from statistically towards systematically dominated measurements. To sustain the 
concept of output classes we further enforce this ansatz through an explicit 
constraint in \LSANNT, retaining the result of the pre-training \CEmin resembling 
a good separation between classes. For the subsequent \SANNT, we modify \LSANNT to 
read
  \begin{equation}
    \label{eq:multiclass-sannt-loss}
    \LSANNT^{\mathrm{mult.}} = \sum\limits_{s}\Drs + \ol\,g(\,\cdot
    \,),
  \end{equation}
with 
\begin{equation*}
    g(\,\cdot\,) = \mathrm{CE'} - \CEmin,
\end{equation*}
following an approach as proposed in Ref.~\cite{MDMM}, with \ol as an additional 
trainable parameter of the NN model. Formally, this turns the \SANNT into an 
optimization task with the achieved separation between classes after pre-training, 
expressed by \CEmin, as a constraint equation, where \ol takes the role of a 
Lagrange multiplier. In the nonexcluded case of $g(\,\cdot\,)<0$, we set $\ol=0$ 
to allow improvements of the process separation during \SANNT. By the sum in 
Eq.~(\ref{eq:multiclass-sannt-loss}), all signal processes $s$ are treated equally, 
which is an arbitrary choice. 

\begin{figure*}[bth]
  \centering
  \includegraphics[width=0.98\textwidth]{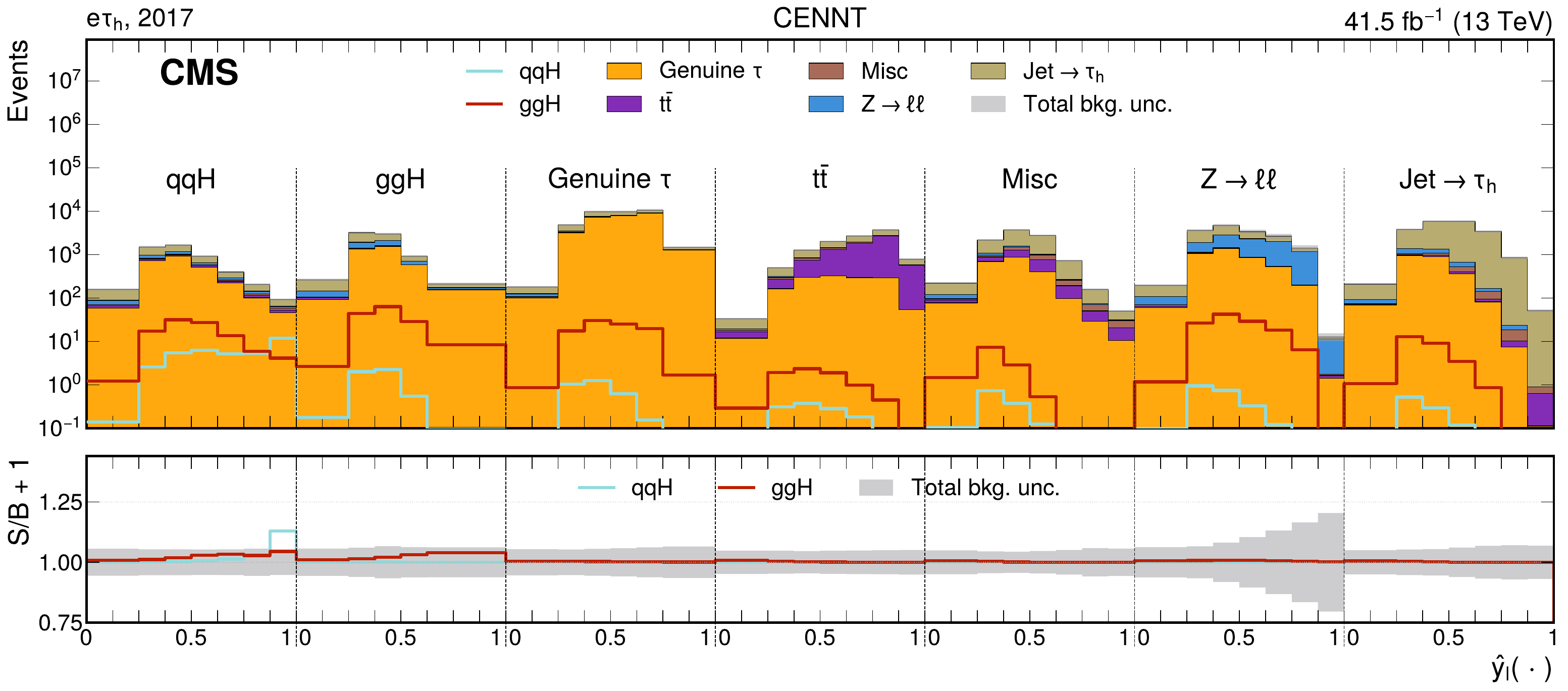}
  \includegraphics[width=0.98\textwidth]{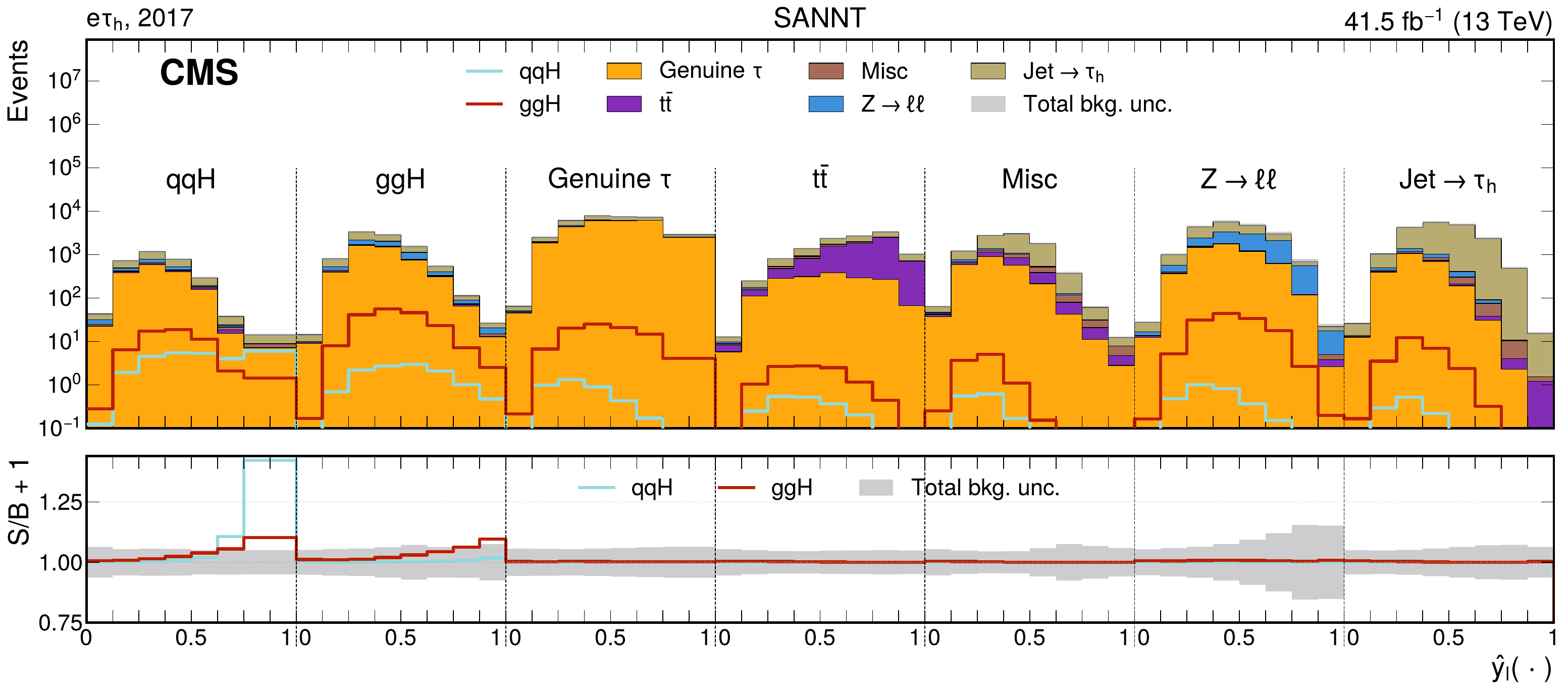}
  \caption{
    Expected distributions of \yl for multiclass classification, based on seven 
    event classes, as used for a differential STXS cross section measurement 
    of \PH production in Ref.~\cite{CMS:2022kdi}, prior to any fit to 
    \Asimov. In the upper (lower) part of the figure the results obtained after 
    \CENNT (\SANNT) are shown. The background processes of \DataModel are indicated 
    by stacked, colored, filled histograms. The expected \ggH and \qqH contributions 
    are indicated by the nonstacked open histograms. In the lower panels of the 
    figure the expected values of $S/B+1$ are shown. The gray bands correspond 
    to the combined statistical and systematic uncertainty in the background 
    model. 
  }\label{fig:multiclass-discriminant}
\end{figure*}

Figure~\ref{fig:multiclass-discriminant} (upper) shows the expected distributions 
of \yl for each background and signal class, and the expected values of $S/B+1$ 
after \CENNT prior to any fit to \Asimov. Individual background processes of 
\DataModel are indicated by stacked, differently colored, filled histograms. The 
expected \ggH and \qqH contributions are indicated by the nonstacked, red- and 
cyan-colored, open histograms. As for the previous discussion in 
Section~\ref{sec:comparison-cennt-sannt}, the gray bands in the lower panels 
of the figure correspond to the combined statistical and systematic uncertainty 
in the background model. The same distributions after \SANNT are shown in 
Fig.~\ref{fig:multiclass-discriminant} (lower). The change of distributions and 
more prominently the improved signal separation, with reduced uncertainty are 
visible from the comparison of the corresponding distributions, indicating the 
success of the training. 

\begin{figure*}[bth]
  \centering
  \includegraphics[width=0.32\textwidth]{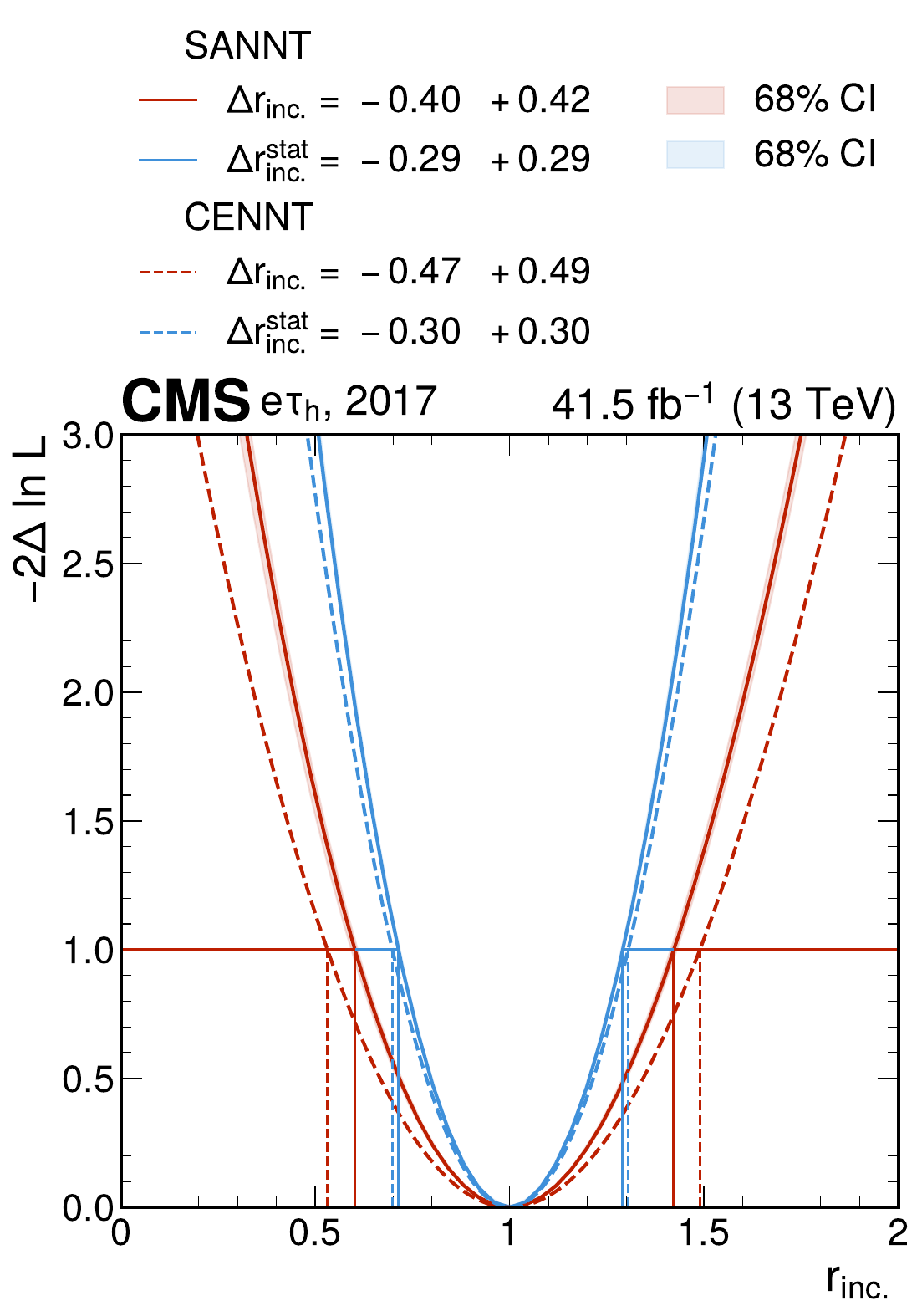}
  \includegraphics[width=0.32\textwidth]{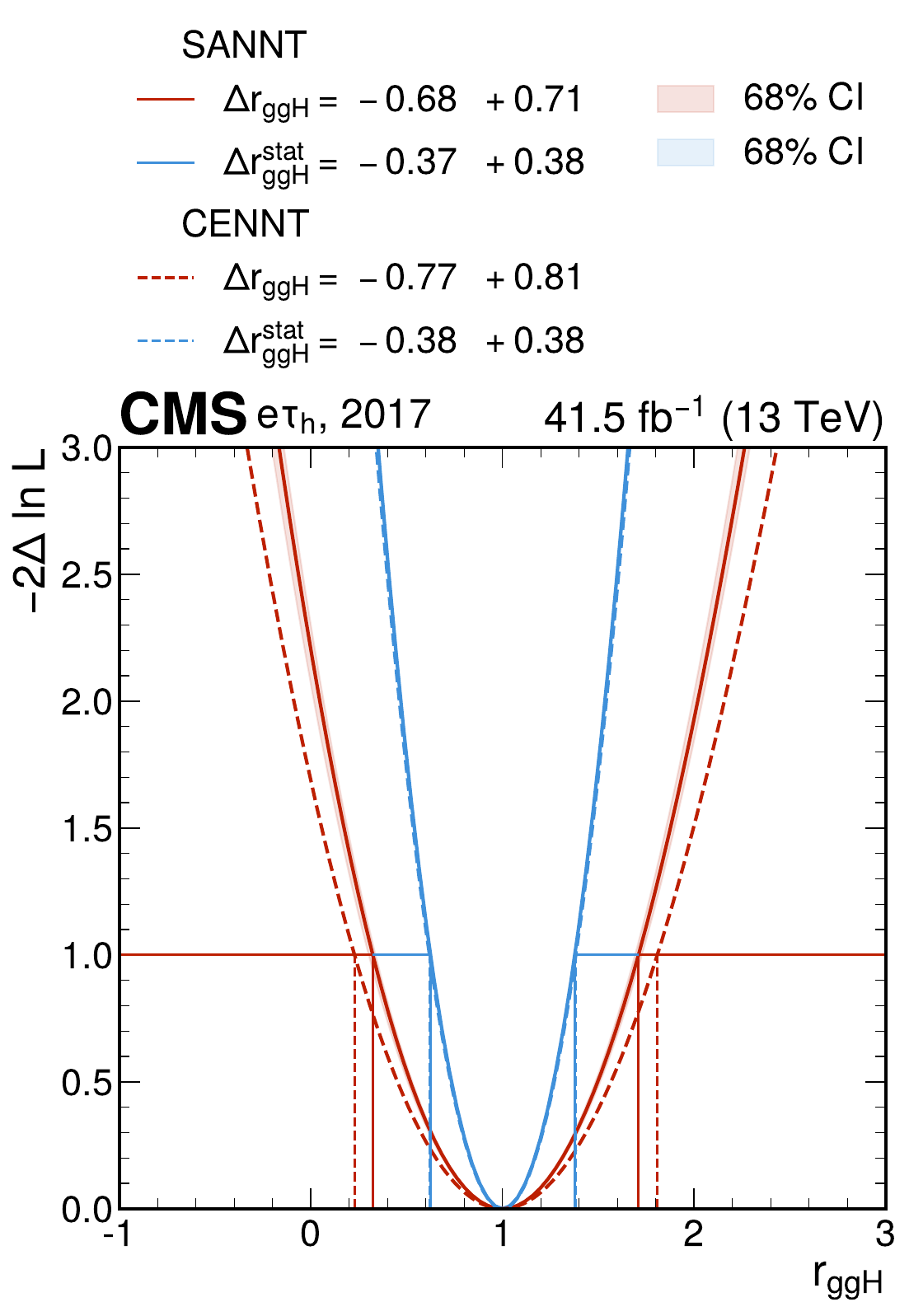}
  \includegraphics[width=0.32\textwidth]{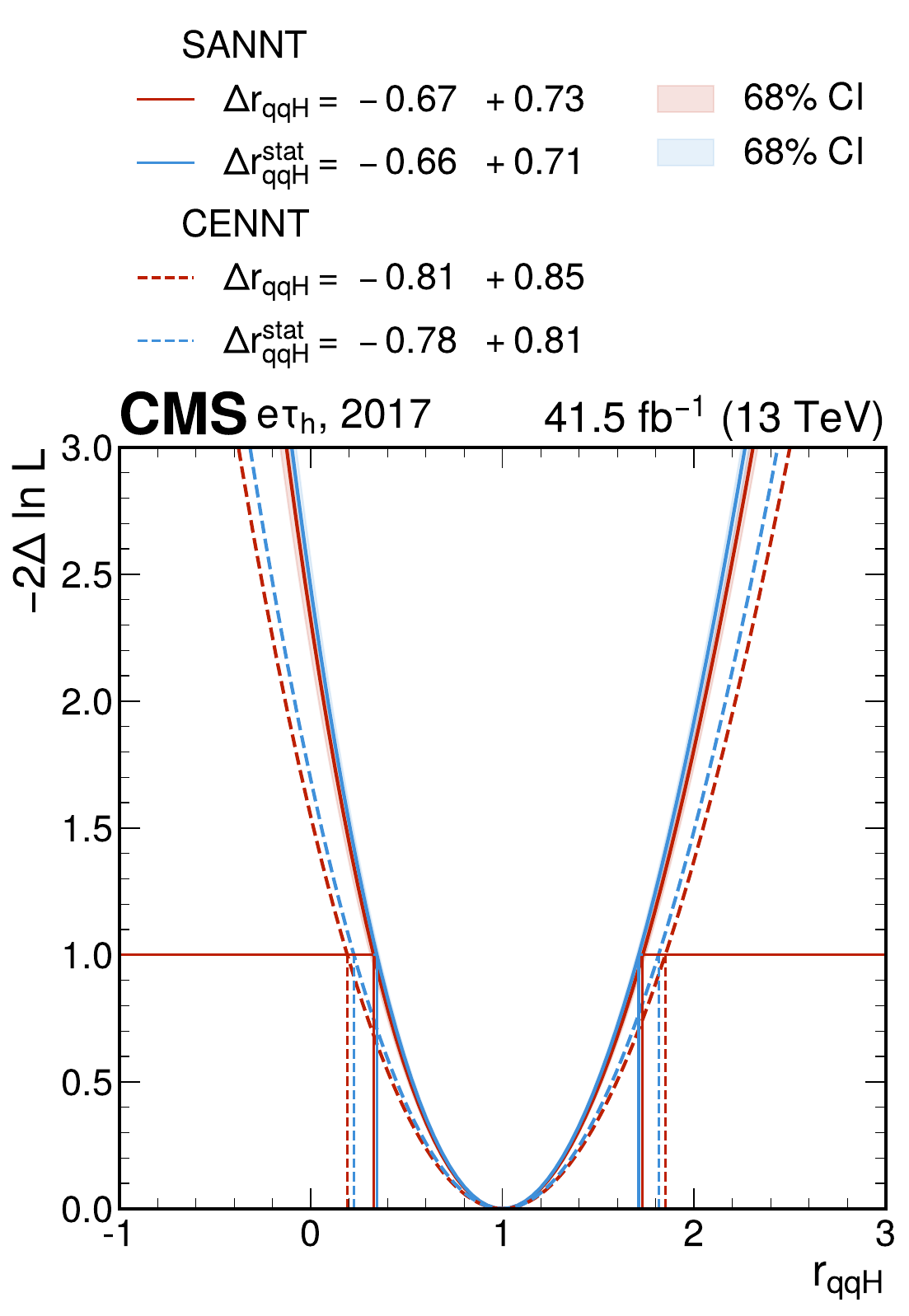}
  \caption{
    Negative log of the profile likelihood \NLL as a function of \rs, for a 
    differential STXS cross section measurement of \PH production in 
    the \HTT decay channel, taking (red) all and (blue) only the statistical 
    uncertainties in \Drs into account. In the left plot \rinc for an inclusive 
    measurement is shown, and in the middle and right plots \rggH and \rqqH for 
    a combined differential STXS measurement of these two contributions to the 
    signal in two bins are shown. The results as obtained from the \CENNT are 
    indicated by the dashed lines, and the median expected results of ensembles 
    of 100 repetitions of the \SANNT varying random initializations are indicated 
    by the continuous lines. The red and blue shaded bands surrounding the median 
    expectations indicate 68\% confidence intervals (CI) from these ensembles.
  }\label{fig:multiclass-NLL-scan}
\end{figure*}

Figure~\ref{fig:multiclass-NLL-scan} shows distributions of \NLL, based on the 
input distributions as depicted in Fig.~\ref{fig:multiclass-discriminant}, as a 
function of \rs. As in the previous discussion, in Section~\ref{sec:comparison-cennt-sannt}, 
for the red (blue) curves all (only the statistical) uncertainties in \Drs have 
been taken into account. In the left part of the figure \rinc, for an inclusive 
measurement of the \PH production cross section in the \HTT decay channel is 
shown, in the middle and right parts of the figure \rggH and \rqqH for a combined 
differential STXS measurement of these contributions to the signal in two bins, 
are shown. The numerical results of the fits are summarized in 
Table~\ref{tab:multiclass-results}.

\begin{table}[tbh]
  \centering
  \topcaption{
    Expected combined statistical and systematic uncertainties \Drs and statistical 
    uncertainties \DrsStat in the parameters \rinc for an inclusive, and \rggH and 
    \rqqH for a differential STXS, cross section measurement of \PH production 
    in the \HTT decay channel, as obtained from fits to \Asimov. In the second 
    (third) column the results after \SANNT (\CENNT) are shown.  
  }
  \renewcommand{\arraystretch}{1.3}
  \begin{tabular}{lcc}
    & \multicolumn{2}{c}{$\Drs$ $(\DrsStat)$} \\
    \rs & \SANNT & \CENNT \\
    \hline 
    \rinc & $^{+0.42}_{-0.40}$ $\left(\pm0.29\right)$ 
    & $^{+0.49}_{-0.47}$ $\left(\pm0.30\right)$ \\ [\cmsTabSkip]
    \rggH & $^{+0.71}_{-0.68}$ $(^{+0.38}_{-0.37})$ 
    & $^{+0.81}_{-0.77}$ $\left(\pm0.38\right)$ \\ [\cmsTabSkip] 
    \rqqH & $^{+0.73}_{-0.67}$ $(^{+0.71}_{-0.66})$ 
    & $^{+0.85}_{-0.81}$ $\left(^{+0.81}_{-0.78}\right)$ \\ [\cmsTabSkip]
    \hline
  \end{tabular}\label{tab:multiclass-results}
\end{table}

As observed previously, the method is unbiased.  After \SANNT, an improvement of 
15\% relative to the result as obtained after \CENNT is observed for \rinc. For 
\rggH and \rqqH, improvements of 12 and 16\% are observed, respectively. This 
constitutes a significant improvement of the \SANNT over the \CENNT. It is the 
first time that a multiclass-classification \SANNT has been successfully 
demonstrated. 

We observe that \SANNT leads to an improvement of the statistical component of 
\rqqH, where the contribution of systematic variations to the overall uncertainty 
has been small, already from the beginning. This can be explained by the fact that 
the targets of both trainings are not in sync. While \LSANNT has been constructed 
to be as close to the measurement target as possible, the target of \CENNT is the 
best possible separation of all processes comprising two signal and five background 
processes, in this case. While for the case of binary classification discussed in 
Section~\ref{sec:comparison-cennt-sannt}, this particular difference between the 
training methods was covered by the experimental setup, it becomes more obvious 
here and emphasizes the conceptual advantage of \SANNT over \CENNT.   

\section{Summary}\label{sec:summary}

We have proposed a neural network training method capable of accounting for the 
effects of systematic variations of the data model used in the training
process and described its extension to neural network multiclass classification.
The procedure has been evaluated on the realistic case of the measurement of 
Higgs boson production via gluon fusion and vector boson fusion in the $\Pgt\Pgt$ 
decay channel at the CMS experiment. The neural network output functions are used 
to infer the signal strengths for inclusive production of Higgs bosons as well as 
for their production via gluon fusion and vector boson fusion. We observe 
improvements of 12 and 16\% in the uncertainty in the signal strengths for gluon 
and vector-boson fusion, respectively, compared with a conventional neural network 
training based on cross-entropy. This is the first time that a neural network 
training capable of accounting for the effects of systematic variations has been 
demonstrated on a data model of this complexity and the first time that such a 
training has been applied to multiclass classification.

\begin{acknowledgments}
  We congratulate our colleagues in the CERN accelerator departments for the excellent performance of the LHC and thank the technical and administrative staffs at CERN and at other CMS institutes for their contributions to the success of the CMS effort. In addition, we gratefully acknowledge the computing centers and personnel of the Worldwide LHC Computing Grid and other centers for delivering so effectively the computing infrastructure essential to our analyses. Finally, we acknowledge the enduring support for the construction and operation of the LHC, the CMS detector, and the supporting computing infrastructure provided by the following funding agencies: SC (Armenia), BMBWF and FWF (Austria); FNRS and FWO (Belgium); CNPq, CAPES, FAPERJ, FAPERGS, and FAPESP (Brazil); MES and BNSF (Bulgaria); CERN; CAS, MoST, and NSFC (China); MINCIENCIAS (Colombia); MSES and CSF (Croatia); RIF (Cyprus); SENESCYT (Ecuador); ERC PRG, RVTT3 and MoER TK202 (Estonia); Academy of Finland, MEC, and HIP (Finland); CEA and CNRS/IN2P3 (France); SRNSF (Georgia); BMBF, DFG, and HGF (Germany); GSRI (Greece); NKFIH (Hungary); DAE and DST (India); IPM (Iran); SFI (Ireland); INFN (Italy); MSIP and NRF (Republic of Korea); MES (Latvia); LMTLT (Lithuania); MOE and UM (Malaysia); BUAP, CINVESTAV, CONACYT, LNS, SEP, and UASLP-FAI (Mexico); MOS (Montenegro); MBIE (New Zealand); PAEC (Pakistan); MES and NSC (Poland); FCT (Portugal); MESTD (Serbia); MICIU/AEI and PCTI (Spain); MOSTR (Sri Lanka); Swiss Funding Agencies (Switzerland); MST (Taipei); MHESI and NSTDA (Thailand); TUBITAK and TENMAK (Turkey); NASU (Ukraine); STFC (United Kingdom); DOE and NSF (USA).

  \hyphenation{Rachada-pisek} Individuals have received support from the Marie-Curie program and the European Research Council and Horizon 2020 Grant, contract Nos.\ 675440, 724704, 752730, 758316, 765710, 824093, 101115353, 101002207, and COST Action CA16108 (European Union); the Leventis Foundation; the Alfred P.\ Sloan Foundation; the Alexander von Humboldt Foundation; the Science Committee, project no. 22rl-037 (Armenia); the Fonds pour la Formation \`a la Recherche dans l'Industrie et dans l'Agriculture (FRIA-Belgium); the Beijing Municipal Science \& Technology Commission, No. Z191100007219010 and Fundamental Research Funds for the Central Universities (China); the Ministry of Education, Youth and Sports (MEYS) of the Czech Republic; the Shota Rustaveli National Science Foundation, grant FR-22-985 (Georgia); the Deutsche Forschungsgemeinschaft (DFG), among others, under Germany's Excellence Strategy -- EXC 2121 ``Quantum Universe" -- 390833306, and under project number 400140256 - GRK2497; the Hellenic Foundation for Research and Innovation (HFRI), Project Number 2288 (Greece); the Hungarian Academy of Sciences, the New National Excellence Program - \'UNKP, the NKFIH research grants K 131991, K 133046, K 138136, K 143460, K 143477, K 146913, K 146914, K 147048, 2020-2.2.1-ED-2021-00181, TKP2021-NKTA-64, and 2021-4.1.2-NEMZ\_KI-2024-00036 (Hungary); the Council of Science and Industrial Research, India; ICSC -- National Research Center for High Performance Computing, Big Data and Quantum Computing and FAIR -- Future Artificial Intelligence Research, funded by the NextGenerationEU program (Italy); the Latvian Council of Science; the Ministry of Education and Science, project no. 2022/WK/14, and the National Science Center, contracts Opus 2021/41/B/ST2/01369 and 2021/43/B/ST2/01552 (Poland); the Funda\c{c}\~ao para a Ci\^encia e a Tecnologia, grant CEECIND/01334/2018 (Portugal); the National Priorities Research Program by Qatar National Research Fund; MICIU/AEI/10.13039/501100011033, ERDF/EU, "European Union NextGenerationEU/PRTR", and Programa Severo Ochoa del Principado de Asturias (Spain); the Chulalongkorn Academic into Its 2nd Century Project Advancement Project, and the National Science, Research and Innovation Fund via the Program Management Unit for Human Resources \& Institutional Development, Research and Innovation, grant B39G670016 (Thailand); the Kavli Foundation; the Nvidia Corporation; the SuperMicro Corporation; the Welch Foundation, contract C-1845; and the Weston Havens Foundation (USA).  
\end{acknowledgments}

\bibliography{auto_generated} 

\clearpage
\numberwithin{figure}{section}
\appendix
\section{Custom backpropagation based on identity operations}\label{sec:appendix}

\begin{figure}[thb!]
  \centering
  \includegraphics[width=0.48\textwidth]{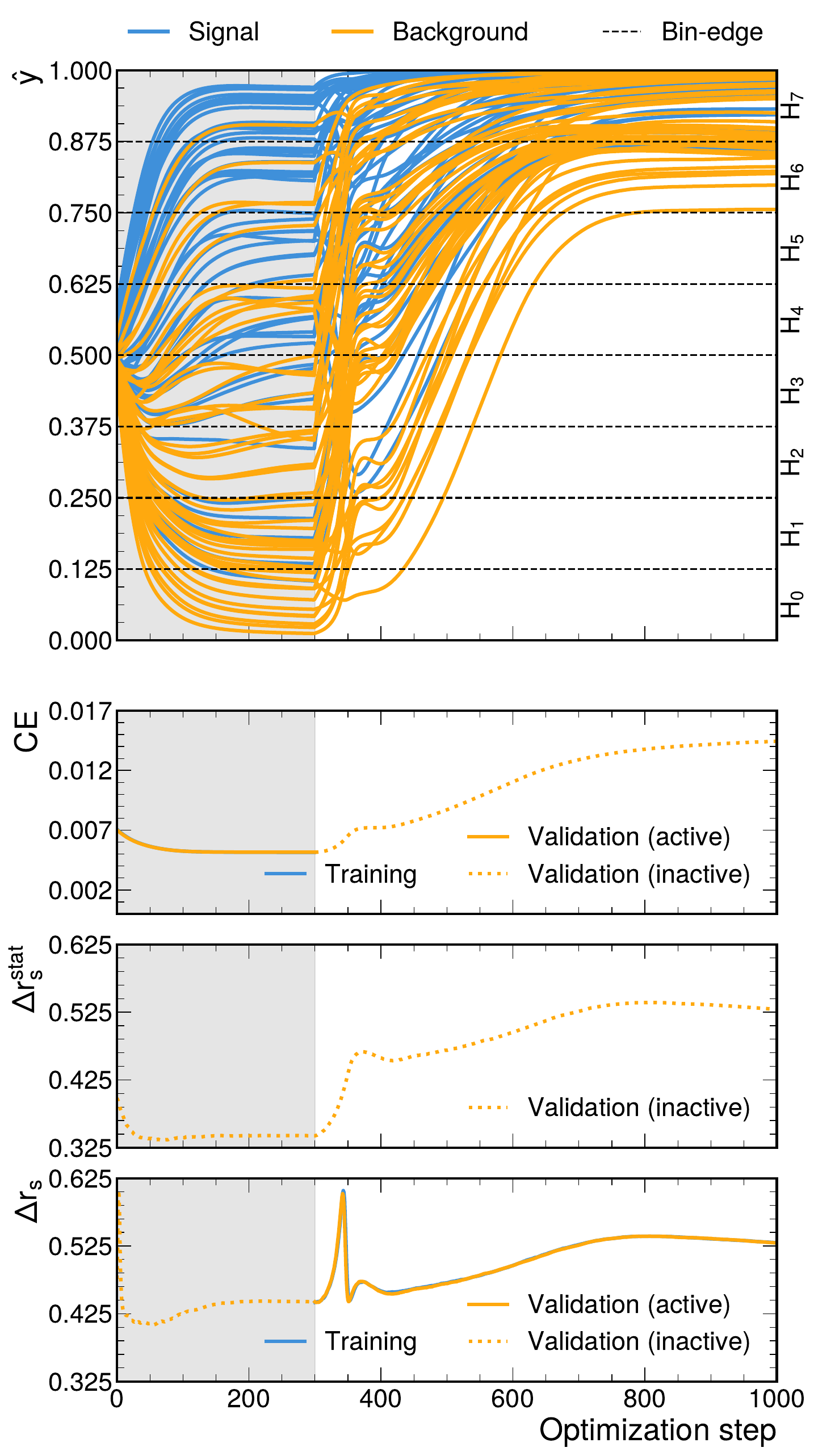}
  \caption{
    Evolution of the loss functions CE, \DrsStat, and \Drs, as used for this 
    paper. Instead of the custom functions \Bprop the identity operation 
    (the so-called straight-through estimator) is used for \SANNT. In the upper 
    panel, the evolution of $\hat{y}$ for randomly selected 50 (blue) signal 
    and 50 (orange) background samples during training is shown. The gray shaded 
    area indicates the pre-training. In the second panel from above the evolution 
    of CE is shown. Though not actively used for the \SANNT \DrsStat is also 
    shown, in the third panel from above. In the lower panel, the evolution of 
    $\Drs$ is shown. The evaluation on the training (validation) data set is 
    indicated in blue (orange). The evolution of inactive loss functions, 
    evaluated on the validation data set, is indicated by the dashed orange 
    curves.
  }\label{fig:ste-training}
\end{figure}

Figure.~\ref{fig:ste-training} shows the evolution of \y, CE, \DrsStat, and \Drs, 
as used for this paper, for a \SANNT where the custom functions \Bprop have been 
replaced by an identity operation, also referred to as STE in the 
literature~\cite{STE}. 

The upper panel of the figure shows the evolution of \y during training, for 
randomly selected 50 (blue) signal and 50 (orange) background samples, which we 
assume to be representative of the full background and signal samples. The 
optimization steps of the training are shown on the $x$ axis and the values of \y 
on the $y$ axis of the figure. The horizontal dashed lines indicate the bin edges 
of $H$, the gray shaded area indicates the pre-training based on CE. We observe 
an even more pronounced collapse of \y, into bins $H_{6}$ and $H_{7}$ in that 
case, as for the original choice of \Bprop, shown in Fig.~\ref{fig:pretraining} 
(left).  

The lower part of the figure shows the evolution of (second panel) CE, (third panel) 
\DrsStat, and (last panel) \Drs over the same period of optimization steps. The 
evaluations on the training and validation samples are indicated by the blue and 
orange curves, respectively. The continuous lines indicate the active loss function 
during each corresponding optimization step. The evolution of inactive loss 
functions on the validation sample is indicated by orange dashed lines. The 
progression of \y and CE during pre-training is identical to 
Fig.~\ref{fig:pretraining} (right), while we have adapted the $y$ axis range here 
to be able to display the full development of CE up to optimization step 1000. 
After pre-training we observe that both \Drs and \DrsStat increase with a pronounced 
structure in the course of \Drs around optimization step 350, coinciding with the 
period of a few optimization steps, during which \y collapses into $H_{6}$ and 
$H_{7}$. Even after 1000 optimization steps, \Drs has not improved beyond its value 
immediately after pre-training, which demonstrates that \Bprop can not be replaced 
by a naive STE and that this change rather destabilizes the training, compromising 
its success.

\cleardoublepage \section{The CMS Collaboration \label{app:collab}}\begin{sloppypar}\hyphenpenalty=5000\widowpenalty=500\clubpenalty=5000
\cmsinstitute{Yerevan Physics Institute, Yerevan, Armenia}
{\tolerance=6000
V.~Chekhovsky, A.~Hayrapetyan, V.~Makarenko\cmsorcid{0000-0002-8406-8605}, A.~Tumasyan\cmsAuthorMark{1}\cmsorcid{0009-0000-0684-6742}
\par}
\cmsinstitute{Institut f\"{u}r Hochenergiephysik, Vienna, Austria}
{\tolerance=6000
W.~Adam\cmsorcid{0000-0001-9099-4341}, J.W.~Andrejkovic, L.~Benato\cmsorcid{0000-0001-5135-7489}, T.~Bergauer\cmsorcid{0000-0002-5786-0293}, S.~Chatterjee\cmsorcid{0000-0003-2660-0349}, K.~Damanakis\cmsorcid{0000-0001-5389-2872}, M.~Dragicevic\cmsorcid{0000-0003-1967-6783}, P.S.~Hussain\cmsorcid{0000-0002-4825-5278}, M.~Jeitler\cmsAuthorMark{2}\cmsorcid{0000-0002-5141-9560}, N.~Krammer\cmsorcid{0000-0002-0548-0985}, A.~Li\cmsorcid{0000-0002-4547-116X}, D.~Liko\cmsorcid{0000-0002-3380-473X}, I.~Mikulec\cmsorcid{0000-0003-0385-2746}, J.~Schieck\cmsAuthorMark{2}\cmsorcid{0000-0002-1058-8093}, R.~Sch\"{o}fbeck\cmsAuthorMark{2}\cmsorcid{0000-0002-2332-8784}, D.~Schwarz\cmsorcid{0000-0002-3821-7331}, M.~Sonawane\cmsorcid{0000-0003-0510-7010}, W.~Waltenberger\cmsorcid{0000-0002-6215-7228}, C.-E.~Wulz\cmsAuthorMark{2}\cmsorcid{0000-0001-9226-5812}
% ...
\par}
\vskip\cmsinstskip
\dag:~Deceased\\
$^{1}$Also at Yerevan State University, Yerevan, Armenia\\
$^{2}$Also at TU Wien, Vienna, Austria\\
$^{3}$Also at Ghent University, Ghent, Belgium\\
$^{4}$Also at Universidade do Estado do Rio de Janeiro, Rio de Janeiro, Brazil\\
$^{5}$Also at FACAMP - Faculdades de Campinas, Sao Paulo, Brazil\\
$^{6}$Also at Universidade Estadual de Campinas, Campinas, Brazil\\
$^{7}$Also at Federal University of Rio Grande do Sul, Porto Alegre, Brazil\\
$^{8}$Also at University of Chinese Academy of Sciences, Beijing, China\\
$^{9}$Also at China Center of Advanced Science and Technology, Beijing, China\\
$^{10}$Also at University of Chinese Academy of Sciences, Beijing, China\\
$^{11}$Also at China Spallation Neutron Source, Guangdong, China\\
$^{12}$Now at Henan Normal University, Xinxiang, China\\
$^{13}$Also at University of Shanghai for Science and Technology, Shanghai, China\\
$^{14}$Now at The University of Iowa, Iowa City, Iowa, USA\\
$^{15}$Also at an institute formerly covered by a cooperation agreement with CERN\\
$^{16}$Also at Cairo University, Cairo, Egypt\\
$^{17}$Also at Suez University, Suez, Egypt\\
$^{18}$Now at British University in Egypt, Cairo, Egypt\\
$^{19}$Also at Purdue University, West Lafayette, Indiana, USA\\
$^{20}$Also at Universit\'{e} de Haute Alsace, Mulhouse, France\\
$^{21}$Also at Istinye University, Istanbul, Turkey\\
$^{22}$Also at an international laboratory covered by a cooperation agreement with CERN\\
$^{23}$Also at The University of the State of Amazonas, Manaus, Brazil\\
$^{24}$Also at University of Hamburg, Hamburg, Germany\\
$^{25}$Also at RWTH Aachen University, III. Physikalisches Institut A, Aachen, Germany\\
$^{26}$Also at Bergische University Wuppertal (BUW), Wuppertal, Germany\\
$^{27}$Also at Brandenburg University of Technology, Cottbus, Germany\\
$^{28}$Also at Forschungszentrum J\"{u}lich, Juelich, Germany\\
$^{29}$Also at CERN, European Organization for Nuclear Research, Geneva, Switzerland\\
$^{30}$Also at HUN-REN ATOMKI - Institute of Nuclear Research, Debrecen, Hungary\\
$^{31}$Now at Universitatea Babes-Bolyai - Facultatea de Fizica, Cluj-Napoca, Romania\\
$^{32}$Also at MTA-ELTE Lend\"{u}let CMS Particle and Nuclear Physics Group, E\"{o}tv\"{o}s Lor\'{a}nd University, Budapest, Hungary\\
$^{33}$Also at HUN-REN Wigner Research Centre for Physics, Budapest, Hungary\\
$^{34}$Also at Physics Department, Faculty of Science, Assiut University, Assiut, Egypt\\
$^{35}$Also at Punjab Agricultural University, Ludhiana, India\\
$^{36}$Also at University of Visva-Bharati, Santiniketan, India\\
$^{37}$Also at Indian Institute of Science (IISc), Bangalore, India\\
$^{38}$Also at Amity University Uttar Pradesh, Noida, India\\
$^{39}$Also at IIT Bhubaneswar, Bhubaneswar, India\\
$^{40}$Also at Institute of Physics, Bhubaneswar, India\\
$^{41}$Also at University of Hyderabad, Hyderabad, India\\
$^{42}$Also at Deutsches Elektronen-Synchrotron, Hamburg, Germany\\
$^{43}$Also at Isfahan University of Technology, Isfahan, Iran\\
$^{44}$Also at Sharif University of Technology, Tehran, Iran\\
$^{45}$Also at Department of Physics, University of Science and Technology of Mazandaran, Behshahr, Iran\\
$^{46}$Also at Department of Physics, Faculty of Science, Arak University, ARAK, Iran\\
$^{47}$Also at Helwan University, Cairo, Egypt\\
$^{48}$Also at Italian National Agency for New Technologies, Energy and Sustainable Economic Development, Bologna, Italy\\
$^{49}$Also at Centro Siciliano di Fisica Nucleare e di Struttura Della Materia, Catania, Italy\\
$^{50}$Also at Universit\`{a} degli Studi Guglielmo Marconi, Roma, Italy\\
$^{51}$Also at Scuola Superiore Meridionale, Universit\`{a} di Napoli 'Federico II', Napoli, Italy\\
$^{52}$Also at Fermi National Accelerator Laboratory, Batavia, Illinois, USA\\
$^{53}$Also at Lulea University of Technology, Lulea, Sweden\\
$^{54}$Also at Laboratori Nazionali di Legnaro dell'INFN, Legnaro, Italy\\
$^{55}$Also at Consiglio Nazionale delle Ricerche - Istituto Officina dei Materiali, Perugia, Italy\\
$^{56}$Also at Institut de Physique des 2 Infinis de Lyon (IP2I ), Villeurbanne, France\\
$^{57}$Also at Department of Applied Physics, Faculty of Science and Technology, Universiti Kebangsaan Malaysia, Bangi, Malaysia\\
$^{58}$Also at Consejo Nacional de Ciencia y Tecnolog\'{i}a, Mexico City, Mexico\\
$^{59}$Also at Trincomalee Campus, Eastern University, Sri Lanka, Nilaveli, Sri Lanka\\
$^{60}$Also at Saegis Campus, Nugegoda, Sri Lanka\\
$^{61}$Also at National and Kapodistrian University of Athens, Athens, Greece\\
$^{62}$Also at Ecole Polytechnique F\'{e}d\'{e}rale Lausanne, Lausanne, Switzerland\\
$^{63}$Also at Universit\"{a}t Z\"{u}rich, Zurich, Switzerland\\
$^{64}$Also at Stefan Meyer Institute for Subatomic Physics, Vienna, Austria\\
$^{65}$Also at Laboratoire d'Annecy-le-Vieux de Physique des Particules, IN2P3-CNRS, Annecy-le-Vieux, France\\
$^{66}$Also at Near East University, Research Center of Experimental Health Science, Mersin, Turkey\\
$^{67}$Also at Konya Technical University, Konya, Turkey\\
$^{68}$Also at Izmir Bakircay University, Izmir, Turkey\\
$^{69}$Also at Adiyaman University, Adiyaman, Turkey\\
$^{70}$Also at Bozok Universitetesi Rekt\"{o}rl\"{u}g\"{u}, Yozgat, Turkey\\
$^{71}$Also at Marmara University, Istanbul, Turkey\\
$^{72}$Also at Milli Savunma University, Istanbul, Turkey\\
$^{73}$Also at Kafkas University, Kars, Turkey\\
$^{74}$Now at Istanbul Okan University, Istanbul, Turkey\\
$^{75}$Also at Hacettepe University, Ankara, Turkey\\
$^{76}$Also at Erzincan Binali Yildirim University, Erzincan, Turkey\\
$^{77}$Also at Istanbul University -  Cerrahpasa, Faculty of Engineering, Istanbul, Turkey\\
$^{78}$Also at Yildiz Technical University, Istanbul, Turkey\\
$^{79}$Also at School of Physics and Astronomy, University of Southampton, Southampton, United Kingdom\\
$^{80}$Also at IPPP Durham University, Durham, United Kingdom\\
$^{81}$Also at Monash University, Faculty of Science, Clayton, Australia\\
$^{82}$Also at Universit\`{a} di Torino, Torino, Italy\\
$^{83}$Also at Bethel University, St. Paul, Minnesota, USA\\
$^{84}$Also at Karamano\u {g}lu Mehmetbey University, Karaman, Turkey\\
$^{85}$Also at California Institute of Technology, Pasadena, California, USA\\
$^{86}$Also at United States Naval Academy, Annapolis, Maryland, USA\\
$^{87}$Also at Ain Shams University, Cairo, Egypt\\
$^{88}$Also at Bingol University, Bingol, Turkey\\
$^{89}$Also at Georgian Technical University, Tbilisi, Georgia\\
$^{90}$Also at Sinop University, Sinop, Turkey\\
$^{91}$Also at Erciyes University, Kayseri, Turkey\\
$^{92}$Also at Horia Hulubei National Institute of Physics and Nuclear Engineering (IFIN-HH), Bucharest, Romania\\
$^{93}$Now at another institute formerly covered by a cooperation agreement with CERN\\
$^{94}$Also at Texas A\&M University at Qatar, Doha, Qatar\\
$^{95}$Also at Kyungpook National University, Daegu, Korea\\
$^{96}$Also at Yerevan Physics Institute, Yerevan, Armenia\\
$^{97}$Also at another international laboratory covered by a cooperation agreement with CERN\\
$^{98}$Also at Imperial College, London, United Kingdom\\
$^{99}$Also at Institute of Nuclear Physics of the Uzbekistan Academy of Sciences, Tashkent, Uzbekistan\\
$^{100}$Also at another institute formerly covered by a cooperation agreement with CERN\\
\end{sloppypar}
\end{document}